\documentclass[aps, prd, amsmath, amssymb, twocolumn, 10pt, a4paper, superscriptaddress,nofootinbib,longbibliography]{revtex4-2}

\usepackage[percent]{overpic}
\usepackage{amsmath}
\usepackage{multirow}
\usepackage{booktabs}
\usepackage{array}
\usepackage{comment}
\usepackage{amssymb}
\usepackage{mathrsfs}
\usepackage{xcolor}
\usepackage{colortbl}
\usepackage{scalerel}
\usepackage{graphicx}
\usepackage{rotating}
\usepackage{float}
\usepackage{orcidlink}
\usepackage{rotating, makecell, xcolor} 

\usepackage{adjustbox}

\newcommand{\ie}{i.e.,~}
\newcommand{\mytable}[2]{
  \renewcommand{\arraystretch}{1.0}
  \begin{tabular}{#1}
    #2
  \end{tabular}
}

\usepackage{hyperref}
\hypersetup{colorlinks=true, linkcolor=magenta, citecolor=magenta, filecolor=blue, urlcolor=magenta}

\begin{document}

\title{Relativistic Spherical Accretion in Nonrotating Black Hole Spacetimes: \\
Astrophysical Implications}

\author{David Barrero-González\,\orcidlink{0000-0003-0665-3320}}
\email{dbarrero@astro.unam.mx}
\affiliation{Instituto de Astronomía, Universidad Nacional Autónoma de México, AP 70-264, Ciudad de México 04510, México}

\author{Alejandro Cruz-Osorio\,\orcidlink{0000-0002-3945-6342}}
\email{aosorio@astro.unam.mx}
\affiliation{Instituto de Astronomía, Universidad Nacional Autónoma de México, AP 70-264, Ciudad de México 04510, México}

\begin{abstract}
We present a systematic analysis of spherically symmetric accretion flows onto thirteen non-rotating black hole spacetimes, including charged and regular black holes, scalar and dilaton fields, dark energy, cosmological constants, modified gravity, and parametrized metrics. We generalize the Michel solution to arbitrary diagonal spacetimes and investigate the density, temperature, velocity, Mach number, mass accretion rate, and bremsstrahlung luminosity equilibrium profiles. Using the Schwarzschild solution as a reference, we find significant deviations in the accretion flow, with cosmological-constant and dark-energy models strongly suppressing accretion, while regular and Yukawa-like spacetimes generally enhance it. We provide fitting formulas for the accretion rates and luminosities as functions of the spacetime parameters. We then explore two astrophysical applications: accretion onto supermassive black holes, such as Sgr\,A* and M87*, and the inferred progenitor masses of the intermediate-mass black holes associated with the GW190521 gravitational wave event. We find that observational uncertainties may allow us to probe signatures of deviations from the Schwarzschild spacetime.
\end{abstract}
    
\maketitle
    \section{Introduction} \label{sec:intro}
    Spherically symmetric accretion onto a compact object was studied for the first 
    time by \cite{Bondi1952}, deriving a solution within the framework of Newtonian 
    gravity for a spherically symmetric, stationary gas flow onto a compact object 
    such as a star. The model assumes a perfect fluid without magnetic fields. In 
    this scenario, the compact object is embedded in an infinitely large, homogeneous 
    gas cloud with constant density $\rho_\infty$ and pressure $p_\infty$ at infinity. 
    As the gas moves inward toward the compact object, it develops a radial velocity 
    profile, and its temperature, density, and other thermodynamic properties evolve 
    purely under the influence of gravity, resulting on a steady state accretion 
    process.
    This analysis was later extended to black holes by \citet{Michel1972}, considering 
    the same gas properties but within a general relativistic framework, modeling 
    accretion onto a Schwarzschild black hole. In this context, the accretion flow 
    becomes transonic, transitioning from subsonic to supersonic as the gas approaches 
    the event horizon.
    
    The Michel solution has been extended in various directions, including the studies of \cite{Park1998, Chaverra2015, Abbas2021}, which retain spherical symmetry. Its generalization to higher-dimensional black holes was presented by \citet{Anslyn2013}. In \citet{Zanotti2003}, a perturbation of the Michel solution was introduced to model gravitational wave emission. Subsequently, \citet{AguayoOrtiz2021} extended the analysis to Kerr black holes and considered different equations of state. An interesting application was explored by \citet{Richards2021}, who studied a non-rotating black hole embedded within a non-rotating neutron star. More recently, spherical accretion onto boson stars has been investigated by \citet{Passos2025}.
    
    Spherical accretion is particularly valuable as a theoretical benchmark because it admits an exact solution that can be studied with high precision. Nevertheless, accretion onto astrophysical black holes is considerably more complex and generally requires more sophisticated modeling. An important example is the Bondi--Hoyle--Lyttleton (BHL) accretion model, originally introduced by \citet{HoyleLyttleton1939}, which was introduced earlier the spherical accretion model developed by Bondi. In the BHL scenario, a compact object moves through a homogeneous gas, resulting in a flow that is intrinsically non-radial and therefore differs fundamentally from the purely radial velocity profile characteristic of Bondi accretion. Extending these accretion models to black holes introduces further complications associated with relativistic gravity and the geometry of the flow. A fully relativistic treatment of accretion scenario was presented by \citet{Petrich1989}. More recently, this solution has been extended to a variety of astrophysical settings through general relativistic hydrodynamic simulations. These investigations have considered accretion onto both non-rotating and rotating black holes, as well as black holes endowed with scalar hair, while progressively incorporating more realistic features of the accreting medium, such as density and velocity gradients and localized inhomogeneities or clumps \cite{Cruz2012,Cruz2013,LoraClavijo2015,Cruz2016,Cruz2017,Cruz2020b,Cruz2023,Cruz2026b}, as well as the effects of stiff equation of states \cite{Ewell2026}.
    
    Although numerous studies have explored the effects of the equation of state on the properties 
    of accreting gas, relatively limited attention has been given to the influence of modifying 
    the spacetime geometry itself. An example of such work is presented by \citet{Bauer2022}, 
    which compares general relativity with alternative theories of gravity--specifically scalar 
    Gauss-Bonnet gravity and the Rezzolla-Zhidenko parametrized metric. In this study, the 
    authors evaluate the size of the black hole shadow by computing the impact parameter, 
    a quantity that directly influences photon trajectories and, consequently, the apparent size 
    of the photon ring as observed by the EHT collaboration \citep{EHTCollaboration2019,
    EHTCollaboration2022}.
   
    In this work, we generalize the Michel solution to arbitrary non-rotating black hole spacetimes, extending the analysis of spherical accretion to geometries with a broad range of physical properties. We consider, in particular, spacetimes characterized by electric and magnetic charges, scalar and dilaton fields, dark energy contributions, and other deviations from the Schwarzschild geometry. Our goal is to quantify how modifications of the spacetime geometry influence the accretion rate onto supermassive black holes and determine how they affect the inferred progenitor masses of intermediate mass black holes. 
   
    This paper is structured as follows. In Section \ref{sec:Michel}, we describe the extension 
    of the Michel solution to an arbitrary diagonal metric. Section \ref{sec:spacetimemodels} 
    introduces the family of spacetimes considered in this work. In Sections \ref{sec:morphology}, 
    \ref{sec:accretion_rates}, and \ref{sec:bremsstrahlung} we present the results of flow 
    morphology, accretion rates and bremsstrahlung luminosity obtained for spherical accretion 
    in different spacetimes, and in Section \ref{sec:conclusions}, we summarize our conclusions.
    Throughout this paper, we use geometrical units, setting $G=c=1$ and the metric signature ($-,+,+,+$).
    
    \section{Michel solution for a diagonal spacetime} \label{sec:Michel}
    
    Following the formalism introduced by \citet{Michel1972}, we start from the fully covariant conservation equations of general relativistic hydrodynamics:
    
    \begin{align}
    \nabla_\mu J^{\mu}& =0,\label{eq:flux} \\
    \nabla_\mu T^{\mu \nu}& =0,\label{eq:energy}
    \end{align}
    
    where $J^\mu=\rho U^\mu$ is the mass flux, $\rho$ is the gas rest mass density and 
    $U^\mu$ is fluid four-velocity components. Here, we consider a neutral, perfect fluid, 
    neglecting viscosity and magnetic fields described by the stress-energy tensor 
    $T^{\mu \nu}=(P+\tilde{\mu})U^\mu U^\nu+P g^{\mu \nu}$, where $P$ is the gas pressure, 
    $\tilde{\mu} = \rho c^{2} + \epsilon$ is the total energy density, and $\epsilon$ is the 
    internal energy.  The gas pressure and rest-mass density satisfy a polytropic equation 
    of state, $P= \kappa \rho^{\tilde{\gamma}}$, where $\tilde{\gamma}$ and $\kappa$ are the adiabatic 
    index and adiabatic constant, respectively.
    We adopt the mixed-index notation used in Michel's formulation, namely, 
    $\nabla_{\mu} T^{\mu}{}_{\nu}=0$. Here, $\nabla_{\mu}$ denotes the covariant 
    derivative associated with the curved spacetime.
    
    The spacetime line element is expressed in its generic form as $ds^2 = g_{\mu \nu} dx^\mu dx^\nu$, 
    where $g_{\mu \nu}$ is the four-dimensional metric tensor. For non-rotating black hole 
    spacetimes analyzed in this work, $\mu = \nu$, which corresponds to a diagonal metric. 
    Appendix~\ref{app:spacetimes} summarises the spacetimes 
    analyzed in this work, which will be described in detail in the following sections.
    
    Following Michel's solution, we assume stationary and spherically symmetric accretion, 
    such that the fluid four-velocity takes the form $U^{\mu} = (U^{t},\, U^{r},\, 0,\, 0)$. Under 
    these assumptions, the conservation equations \eqref{eq:flux} and \eqref{eq:energy} 
    reduce to the following two differential equations:
     
    \begin{equation}
    \label{conservation_simplified}
    \dfrac{d}{dr}(J^{r}\sqrt{-g})=0, \qquad \dfrac{d}{dr}(T^{r}{}_{t}\sqrt{-g})=0.
    \end{equation}
    Combining the normalization condition $U^{\mu} U_{\mu} = -1$ with Eq.~\eqref{conservation_simplified}, 
    we derive the conserved quantities corresponding to the mass flux and radial momentum:
    
    \begin{equation}
    \label{constantsC1C2}
    C_1=\rho U^r\sqrt{-g}\, ,\qquad
    C_2=(P+\tilde{\mu})U_tU^r\sqrt{-g}\,,
    \end{equation}
 
    from these two integration constants, we construct the following conserved quantity:
    
    \begin{equation}
    C_3=\left(\dfrac{C_2}{C_1}\right)^2=-\left(\dfrac{P+\tilde{\mu}}{\rho}\right)^2g_{tt}\left(1+g_{rr}(U^r)^2\right).
        \label{constantsC3}
    \end{equation}
    
     Taking the total derivative of $\ln C_i$ with respect to $r$, $U^{r}$ and $\rho$, and performing some algebraic manipulations, we derive the master equation governing the steady-state accretion flow:
     
    \begin{align}
    \label{eq:generaleq}
    & \dfrac{dr}{r}\Bigg[\dfrac{r}{\sqrt{-g}}\dfrac{d\sqrt{-g}}{dr}c_s^2-\dfrac{r(U^r)^2}{2[1+g_{rr}(U^r)^2]}\dfrac{dg_{rr}}{dr}-
    \dfrac{r}{2g_{tt}}\dfrac{dg_{tt}}{dr}\Bigg] \nonumber \\
    &+ \dfrac{dU^r}{U^r}\Bigg[c_{s}^{2}-\dfrac{g_{rr}(U^r)^2}{1+g_{rr}(U^r)^2}\Bigg] = 0,
    \end{align}
    
   here, the fluid sound speed, $c_s^2$, is defined in terms of the equation of state as follows:  
     
    \begin{equation}
    c_s^2=\dfrac{d\ln(P+\tilde{\mu})}{d\ln\rho}-1.
    \end{equation}
    The hydrodynamic critical point, $r_{c}$, is defined by the simultaneous vanishing of the two terms in brackets in Eq.~\eqref{eq:generaleq}. Solving these conditions yields the critical radial four-velocity, $U^{r}{}_{c}$, and squared sound speed, $c_{s,c}^{2}$ at critical point $r_{c}$:
    
    \begin{align}
    \label{eq:main_equation}
    (U^r {}_c)^2&=\dfrac{-\dfrac{1}{g_{tt}}\dfrac{dg_{tt}}{dr}}{g_{rr}\left(\dfrac{1}{g_{tt}}\dfrac{dg_{tt}}{dr}-\dfrac{2}{\sqrt{-g}}\dfrac{d\sqrt{-g}}{dr}\right)+\dfrac{dg_{rr}}{dr}}\bigg|_{r_c}, \\
    c_{s,c}^2&=\dfrac{g_{rr}(U^r)^2}{1+g_{rr}(U^r)^2}\bigg|_{r_c}.
    \end{align}
    
The Michel solution can be obtained throughout the radial domain, extending from large radii down to the event horizon. To construct the solution, we use the conserved quantities $C_1$, $C_2$, and $C_3$, given in Eqs.~\eqref{constantsC1C2} and~\eqref{constantsC3}. We express these quantities in terms of the fluid temperature profile, $T(r)$, and determine the temperature at each radial position using the Newton--Raphson iterative method.
Starting from the polytropic equation of state and the definition of the dimensionless gas temperature, $T=P/\rho$, we obtain $\rho = \frac{\rho_c}{T_c^n}T^n$, where $n=1/(\tilde{\gamma}-1)$. Substituting this relation into the conserved quantities and expressing the resulting equations in terms of the temperature and radial four-velocity, we obtain
\begin{align}
C_4 &= T^n U^r \sqrt{-g}\, ,\\
C_5 &= -[1+(1+n)T]^2\,g_{tt} \left[1+g_{rr}(U^r)^2\right].
\end{align}
Here, $C_4$ and $C_5$ are constants determined by the critical-point solution.
These relations allow us to determine the temperature profile, $T(r)$, throughout the radial domain. At a given radial position $r_i$, the radial four-velocity can be expressed in terms of the temperature as
    \begin{align}
    U^r_i&=\dfrac{C_4}{T_i^n \sqrt{-g}}\,, \label{eq:ur_tn}
    \end{align}
We therefore define the root-finding function
    \begin{align}
    f_i&=-[1+(1+n)T_i]^2\,g_{tt}\left[1+g_{rr}(U_i^r)^2\right] - C_5\,,
    \end{align}
The temperature $T_i$ is obtained iteratively using the Newton--Raphson method,
\begin{equation}
T_i^{(k+1)} = T_i^{(k)} -\frac{f_i}{f_i'}, \label{eq:NR}
\end{equation}
where the iteration is continued until the convergence criterion $|f(T_i)/f'(T_i)| < \text{tol}$ is satisfied, with $\mathrm{tol}=10^{-5}$. Once the temperature is determined, the remaining hydrodynamic quantities, including $U_t$, $U^r$, $\rho$, $P$, and $c_s$, can be calculated directly from the corresponding relations.   
Following the solution presented by \citet{RezzollaZanotti2013} for the Schwarzschild black hole, 
we adopt a fiducial model with a rest-mass density of $\rho_c = 0.05$ at the critical radius $r_c = 10M$. Nevertheless, solutions can be constructed for arbitrary values of $\rho_c$ and $r_c$. In order to have directly comparable curves for these radial profiles, we solve the Eq.~\eqref{eq:NR} over the domain $r \in [1.01, 50]\,r_{\rm EH}$, with spatial resolution of $\Delta r= 0.01\, r_{\rm EH} $.
    
    \section{Spacetime solutions}
    \label{sec:spacetimemodels}
    
    A variety of non-rotating black hole spacetimes have been considered in the literature, which can be broadly classified into two main categories:
    (i) non-vacuum solutions within general relativity that incorporate matter or additional 
    fields, such as axions, dilatons, cold dark matter, scalar field dark matter, and electromagnetic 
    fields; and (ii) solutions arising from alternative or modified theories of gravity, in which the
    the gravitational interaction is generalized or modified relative to general relativity.    
    In this work, we consider representative solutions from both categories, restricting our analysis to non-rotating black holes. Most of the spacetimes can be written in the form
    \begin{align}
      ds^2=& -f\,dt^2 + \frac{1}{f}dr^2 + r^2d\Omega^2, \label{eq:sphericalsp}\\
       \text{where} \quad f=& 1 - \frac{2\,M(r)}{r}, \label{eq:efe} \\
       \text{and} \quad d\Omega^2=&d\theta^2+\sin^2\theta d\phi^2. \label{eq:angular_dif}
    \end{align}   
    Note that $M(r)$ represents an effective mass as a function of the radial coordinate, 
    that we will define below for each spacetime. 
    Given the diversity of their intrinsic properties, we further organize these solutions into eight families according to their shared physical characteristics:
    \begin{itemize} 
    \item[\textit{i)}] \textbf{Charged black holes.}
    The Schwarzschild solution assumes that during the gravitational collapse of a star, the 
    resulting black hole retains only mass. However, for a charged compact object, 
    charge conservation suggests that the black hole may retain a nonzero net charge, 
    thereby altering the surrounding spacetime geometry.
    The archetype solution within this family is the Reissner–Nordstr\"om (RN) black hole \citep{Reissner1916,Nordstrom1918}, whose $g_{tt}$ component is given by $f=f(r,Q,P)$:
    \begin{equation}
    f=1-\dfrac{2M}{r}+\dfrac{Q^2+P^2}{r^2},
    \end{equation}
    following equation \eqref{eq:sphericalsp}, where $Q$ and $P$ represent the electric and magnetic charges, respectively.  
    We can distinguish three regimes: $Q^2+P^2<M^2$ describe a black hole with an inner and an outer horizonz, $Q^2+P^2=M^2$ represent an extremal black hole, where there is one degenerate horizon and $Q^2+P^2>M^2$ correspond to a naked singularity, but this is an unphysical scenario. 
    
For a charged, nonrotating black hole, such as the Reissner--Nordström solution, the electromagnetic charge contributes to the spacetime curvature through a term $\propto1/r^2$, and the charge enters linearly into the corresponding field equations. On the other hand, non--abelian charge, such as Yang--Mills charge exhibits nonlinear behavior, so that its contribution to the spacetime curvature appears as an independent term in the metric solution \citep{RinconGomez2024}. Following this previous work, we consider the Yang--Mills (YM) solution for which $f=f(r,Q_{\rm YM},Q)$ is given by
    \begin{equation}
    f=1-\dfrac{2M}{r}+\dfrac{Q^2}{r^2}+Q_{\rm YM},
    \end{equation}

The third spacetime of this class that we consider is the Einstein--Maxwell--dilaton (EMd) black hole. This solution incorporates the contribution of a dilaton field, characterized by the dilaton charge $\Sigma = \frac{P^2-Q^2}{2M}$, where $P$ and $Q$ are the magnetic and electric charges, respectively \cite{Kallosh1992}. The metric function $f=f(r,Q,P)$ is given by
\begin{equation}
	f=\dfrac{(r-r_-)(r-r_+)}{r^2-\Sigma^2},
	\end{equation}
where $r_\pm=M\pm r_0$ and $r_0=M^2+\Sigma^2-P^2-Q^2$.	

    \item[\textit{ii)}] \textbf{Regular black holes.} 
    It is commonly assumed that when a black 
    hole forms, gravitational collapse leads to the creation of a singularity — a region where 
    the spacetime metric becomes undefined and the known laws of physics break down. 
    However, some solutions avoid the formation of a singularity, proposing instead that the 
    metric remains regular throughout the entire spacetime including the point $r=0$. The first 
    model of this kind was introduced by \citet{Bardeen1968}. Later, \citet{ayonbeato2000} 
    provided a mathematical derivation of this solution within the framework of nonlinear 
    electrodynamics. This model has also been extensively employed to test alternative 
    theories of gravity, as discussed in \citet{Bambi2014}. A more general mathematical 
    framework merging these models is presented in \citet{Malafarina2023}, where the three 
    specific solutions considered in this work can be derived from a general formula of an 
    effective mass of the form:
    \begin{equation}
    M(r)=\dfrac{M r^\lambda}{(r^\kappa+q_*^\kappa)^{\lambda/\kappa}},
    \end{equation}
    by setting $\lambda = 3$, $\kappa$ as $\kappa\in\{1,2,3\}$ and $q_*$ as follows: 
    \begin{itemize}
    \item[a)] Hayward solution corresponds to $\kappa=3$ and $q_*=(2\ell^2M)^{1/3}$, where $\ell$ 
    is a length scale, and it represents a small deviation from Schwarzschild spacetime  \citep{Hayward2006}.
    \item[b)]  Classic Bardeen solution \citep{Bardeen1968} is recovered for $\kappa=2$ and $q_*=P$, where $P$ is the magnetic charge.
    \item[c)]  Finally, the case $\kappa = 1$, with $q_* = \ell_S$ corresponds to a deviation 
    from the Schwarzschild solution, the super-Planckian Hair (SPH) spacetime, 
    where the deviations emerge at a 
    characteristic scale $\ell_S\gg\ell_P$, being $\ell_{\rm P}$ the Planck length. This spacetime presents larger deviations from Schwarzschild spacetime than Hayward and Bardeen spacetimes \citep{Cadoni2023}. 
    \end{itemize}
    %
    \item[\textit{iii)}] \textbf{Black holes with cosmological constant.}
    The previous models 
    share the feature of being asymptotically flat spacetimes, \ie far 
    from the black hole, the spacetime approaches the Minkowski solution, where the Ricci 
    scalar is $R=0$, indicating that the geometry is influenced solely by the black hole mass. 
    However, introducing a cosmological constant $\Lambda$ modifies the asymptotic behavior, 
    resulting in $R=4\Lambda$, which implies that $\Lambda$ also affects the spacetime 
    geometry at large scales. This leads to stationary, cosmological-like solutions to Einstein 
    equations. The solution 
    of this family can be obtained by considering an Einstein-Hilbert action of the form:
    
    \begin{equation}
    S=\int d^4x\sqrt{-g}\left(R-2\Lambda\right),
    \end{equation}
    subsequently, the Einstein equations in vacuum are expressed as
    \begin{equation}
    R_{\mu\nu}-\dfrac{1}{2}Rg_{\mu\nu}+\Lambda g_{\mu\nu}=0.
    \end{equation}
    For the spherically symmetric and stationary case, the solution is described by the line element 
    given in equation \eqref{eq:sphericalsp}, where $f=f(r,\Lambda)$ is
    \begin{equation}
    f= 1-\dfrac{2M}{r}-\dfrac{\Lambda r^2}{3}.
    \end{equation}
    This solution, known as the Schwarzschild--de Sitter (SdS) spacetime, was introduced by \citet{Kottler1918} \citep[see also][]{deSitter_1917,Rezzolla2003} and is characterized by a non-asymptotically flat geometry.
As the radial coordinate increases, the curvature of the spacetime -- the universe curvature--remains nonzero but constant.\\
\citet{Yang2023} explored a Schwarzschild-de Sitter spacetime with an antisymmetric Kalb-Ramond tensor field $B_{\mu\nu}$, which introduces breaking of invariance under Lorentz transformations.
 For simplicity, we have adopted the name Lorentz symmetry breaking (LSB) for this spacetime. The $g_{tt}$ component is given by $f=f(r,\lambda_{\rm LSB},\Lambda)$:  
      \begin{equation}
    f=\dfrac{1}{1-\lambda_{\rm LSB}}-\dfrac{2M}{r}-\dfrac{\Lambda r^2}{3(1-\lambda_{\rm LSB})},
    \end{equation}
    where    $\lambda_{\rm LSB}=\xi_2 \langle B_{\mu\nu}\rangle\langle B^{\mu\nu}\rangle$/2, is the  Lorentz-violating parameter and $\xi_2$ is a coupling constant between the Ricci tensor and the Kalb-Ramond field.
 
    \item[\textit{iv)}] \textbf{Naked singularity.} We analyze the Janis-Newman-Winicour (JNW) spacetime  \cite{JanisNewmanWinicour1968}, which describes a compact object surrounded 
    by a massless scalar field of the form:
    \begin{equation}
    \varphi = \dfrac{q}{b}\ln\left(1-\dfrac{b}{r}\right),
    \end{equation}
    where $q$ is the \textit{charge} of the scalar field. The parameter $b$ also obeys $\gamma b=2M$, where $\gamma\in(0,1]$ can be understood as the intensity of the scalar field~\citep{Virbhadra_1997,Sau_2020}. Thus, $b$ is not a fundamental parameter, but a parameter which allows to recover the Schwarzschild solution for $\gamma=1$. The line element is given by
    $ds^2=-f^\gamma  dt^2 +f^{-\gamma}dr^2+f^{1-\gamma}r^2d\Omega^2$ 
    where $f=f(r,\gamma)$ is given by:
    \begin{equation}
    f=1-\dfrac{2M}{\gamma r}.
    \end{equation}
    The Kretschamnn curvature invariant has a factor ${\cal K}\sim (1-2M/\gamma r)^{-4}$, which diverges for any surface $r=2M/\gamma$, for $\gamma<1$, which reflects the fact that the JNW metric does not describe a black hole, but a naked singularity, and this surface represents the limit of spacetime.
    \item[\textit{v)}] \textbf{Traversable Wormhole.}
    A Simpson-Visser (SV) \textit{traversable wormholes} is described by 
    a spacetime capable of connecting either two parallel universes or distant 
    regions within a single universe, as proposed by \citet{Simpson_2019} , The spacetime line element is given by $ds^{2}=-f\,dt^{2}+f^{-1}\,dr^{2}+(r^{2}
    +\tilde a^{2})d\Omega^{2}$, where the metric function $f = f(r, \tilde{a})$ is defined as:
    
    \begin{equation}
    f=1-\dfrac{2M}{\sqrt{r^2+\tilde{a}^2}}, 
    \end{equation}
    here, $\tilde{a} \geq 0$ is a free parameter that allows for the construction 
    of regular black holes and, for specific values, enables the extension of the 
    radial coordinate to negative values ($r < 0$).
     The solution describes different 
    geometries depending on the value of $\tilde{a}$:
    i) the Schwarzschild solution when $\tilde{a} = 0$;
    ii) a regular black hole with a one-way spacelike throat at $r = 0$ for $0 < \tilde{a} < 2M$;
    iii) a one-way wormhole with an extremal null throat at $r = 0$ for $\tilde{a} = 2M$;
    iv) a traversable wormhole when $\tilde{a} > 2M$, in which case no event horizon forms, as $g_{tt} < 0$ throughout all the spatial domain, allowing for two-way travel.
    \item[\textit{vi)}] \textbf{Black holes with dark energy.}
    In \citet{Ratra1988}, a model was introduced in which the accelerated expansion of the Universe arises from a scalar field minimally coupled to gravity, rather than from a cosmological constant. The resulting scalar field dynamics provide a time-dependent source of cosmic acceleration, in contrast to the constant energy density associated with a cosmological constant. Neglecting the contributions from radiation and 
    spatial curvature, the Friedmann equations can be written as 
    $H(z)=H_0 \sqrt{\Omega_m(1+z)^3+\Omega_{\rm DE}(1+z)^{3(1+\omega)}}$ 
    \citep{Peebles1980,DESI2024}, where $H_0$ is the Hubble constant, and $\Omega_m$ and $\Omega_{\rm DE}$ are the dimensionless matter and dark energy density parameters, respectively.   
    Models based on a cosmological constant assume that dark energy remains constant over time, corresponding to $\omega=-1$ \citep[see][]{Peebles1982,Efstathiou1990,Peebles2024}. However, in models where $-1<\omega<-1/3$, known as quintessence, the dark energy density evolves due to the dynamics of the scalar field. Hereafter we will refer to this solution as Dark energy-Hayward (DEH) \cite{LiZhangHuang_2024}, where a 
    non-singular metric of the form given in equation \eqref{eq:sphericalsp} was proposed, 
    with $f=f(r,\ell, \omega)$ encoding the influence of the scalar field:
    \begin{equation}
    f=1-\dfrac{2Mr^2}{r^3+2m\ell^2}+\dfrac{\alpha}{r^{3\omega+1}},
    \end{equation}
    here, $\ell$ represents the same scale parameter used in the Hayward model and 
    $\alpha$ is a normalization constant. We note that $\omega=-1$ yields 
    $f_{\rm DE}=\alpha r^2$, and choosing $\alpha=-\Lambda/3$ recovers a 
    de-Sitter-like spacetime.\\
    
    \item[\textit{vii)}] \textbf{Black holes in $\boldsymbol{f(R)}$ theory.}
    Theories beyond general relativity explored in this work include a spherically 
    symmetric, Yukawa-like solution, which generalizes the Newtonian Yukawa 
    potential. These theories are based on replacing the Ricci scalar $R$ with a 
    function $f(R)$ in the Einstein-Hilbert action. A commonly used example is the 
    Yukawa-like model \citep{Borka2013,DAddio2021}, which is derived from a 
    post-Newtonian approximation and assumes an action of the form:
    \begin{equation}
    S=\int d^4x\sqrt{-g}\bigg[f(R)+\mathcal{X}\mathcal{L}_m\bigg],
    \end{equation}
    where $R$ is the Ricci scalar, $\mathcal{X} = 16\pi G/c^4$, and $\mathcal{L}_m$ 
    is the fluid-matter Lagrangian. An exact solution of the Yukawa-like metric in 
    vacuum was obtained by \citet{DeLaurentis2018} \cite[see also][]{CruzOsorio2021}, 
    where the metric does not satisfy the line element given in equation 
    \eqref{eq:sphericalsp}, but instead is expressed as following:
    
    \begin{equation}
    ds^2    = -[1+\Phi(r)]dt^2 + [1-\Psi(r)]dr^2 + d\Omega^2,
    \end{equation}
    where
    \begin{align}
    \Phi(r) &=-\dfrac{2M}{r}\left(\dfrac{\delta e^{-r/\lambda}+1}{\delta+1}\right),\\
    \Psi(r) &= \dfrac{2M}{r}\left[\dfrac{\delta e^{-r/\lambda}+1}{\delta+1}+\dfrac{\delta r e^{-r/\lambda}/\lambda-2}{\delta+1}\right],
    \end{align}
  where $\lambda$ denotes the spatial scale and $\delta$ is the deviation parameter from general relativity, which is recovered in the limit $\delta=0$.
    \item[\textit{viii)}] \textbf{Parameterized spacetimes.}
    The last family of spacetimes studied 
    in this work corresponds to generic solutions described by an expansion in the $g_{tt}$ 
    and $g_{rr}$ components of the metric, characterized by a set of parameters. Although 
    this solution is more mathematically driven than physically motivated, with the correct set 
    of parameters, it is capable of reproducing any of the known solution families. A key example 
    of these solutions in spherical symmetry is the one introduced by \citet{Rezzolla2014} 
    \cite[for rotating black holes in horizon-penetrating coordinates, see ][]{Ma2024}, 
    the element line describing the solution is the following:
    \begin{equation}
    ds^2=-N^2(r)dt^2+\dfrac{B^2(r)}{N^2(r)}dr^2+r^2d\Omega^2, \label{eq:RZ}
    \end{equation}
    where $N^2(x)=xA(x)$, here $x=1-r_0/r$ and
    $A(x)=1-\epsilon(1-x)+(a_0-\epsilon)(1-x)^2+\tilde A(x)(1-x)^3$, and expansion term $\tilde A(x)$
    is defined as:
      \begin{equation}
    \tilde A(x)=\dfrac{a_1}{1+\dfrac{a_2x}{1+\dfrac{a_3x}{1+\dots}}}.
    \end{equation}
    For simplicity, and following previous analysis, we choose $B^2(r)=1$, and for the 
    deviation parameter respect to a Schwarzschild black hole, $\epsilon=\dfrac{2M}{r_0}-1$, 
    we choose $r_0=2M$ ($\epsilon=0$). In the first order approximation, only two free 
    parameters $a_0$ and $a_1$ are non-zero. So, the Rezzolla-Zhidenko (RZ) metric given in equation 
    \eqref{eq:RZ} reduces to the form of line element given in equation \eqref{eq:sphericalsp}, 
    with $f=N^2$, and $N^2$ given by:
    \end{itemize}
  \begin{equation}
  \hspace{-0.46cm}  
    N^2(r)=\left(1-\dfrac{2M}{r}\right)\left[1+a_0\bigg(\dfrac{2M}{r}\bigg)^2+a_1\bigg(\dfrac{2M}{r}\bigg)^3\right].
    \end{equation}
    We note that any combination of $a_0$ and $a_1$ corresponds to a black hole with an event horizon located at $r=2M$.\\
    It is important to note that the classification presented above is not the only possible 
    one, as some models could belong to more than one category. For example, the Bardeen 
    solution is typically classified as a regular black hole, but due to the magnetic charge 
    component, it can also be categorized among charged black holes. 
    
    In Table~\ref{tab:spacetimes}, we present the metric components, $g_{tt}$ and $g_{rr}$, for each spacetime considered in this work, while Table~\ref{tab:spaceparameters} summarizes the ranges of the corresponding free parameters explored in our analysis.
The parameter ranges were chosen primarily based on the physical constraints of each spacetime, with the aim of excluding configurations containing naked singularities. For the Hayward, Bardeen, and SPH spacetimes, for example, we set the upper limit of the parameter range to the corresponding critical value separating black-hole solutions from configurations with the same causal structure as a flat spacetime, since they represent a compact horizonless object. For the YM and RZ spacetimes, the parameter ranges are instead constrained by EHT observations \citep{Kocherlakota2021b,EHTCollaboration2022}. The third column of Table~\ref{tab:spaceparameters} lists the parameter values that recover the Schwarzschild spacetime, whereas the last column indicates the corresponding values beyond which the solutions develop naked singularities, when is possible.
        
    \section{Flow Morphology}\label{sec:morphology}
    
    \begin{figure*}
\centering
\includegraphics[width=0.75\textwidth]{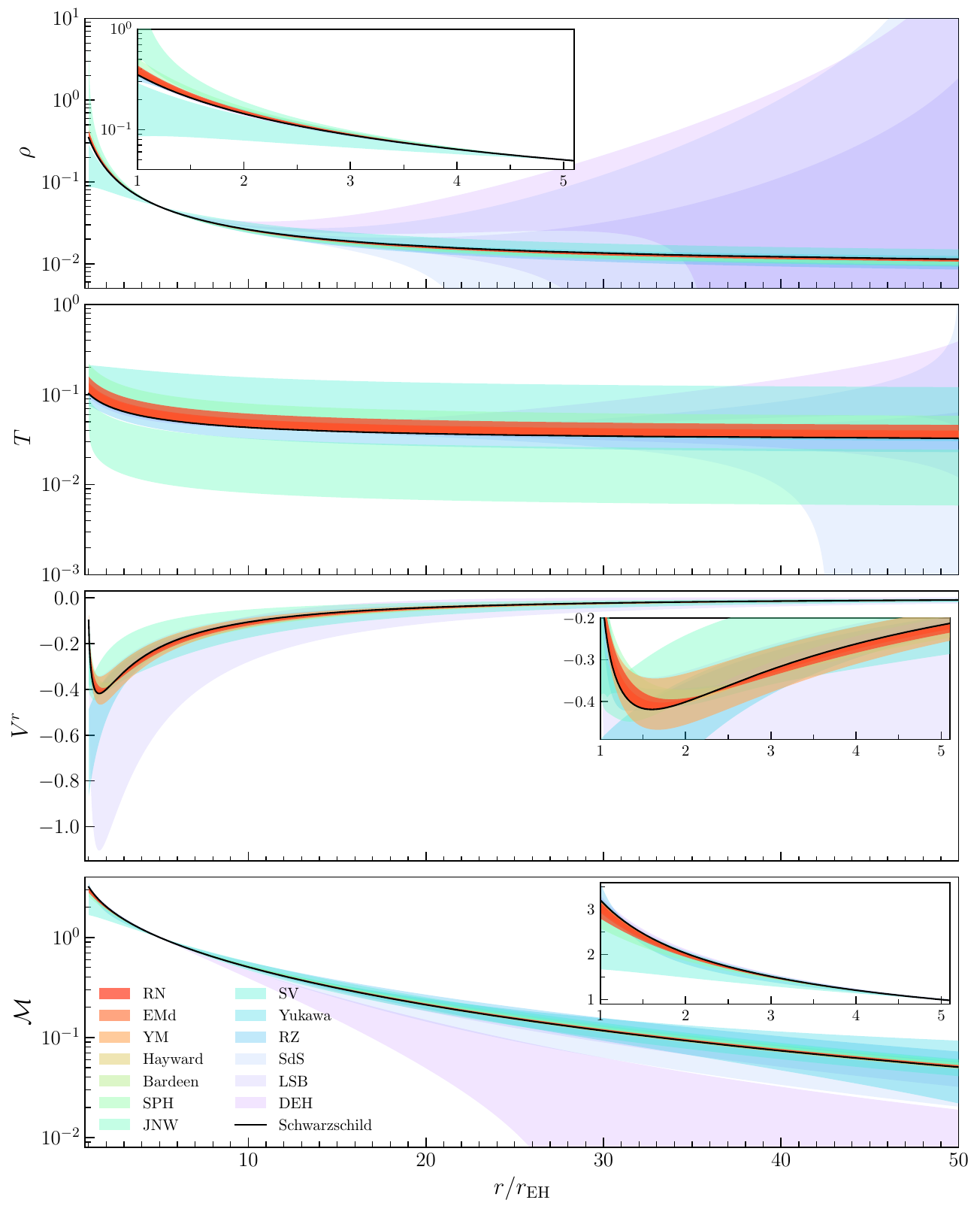}
\caption{Rest-mass density, temperature, radial velocity, and Mach number of the gas as a function of the normalized radius.
The solid line represents Michel solution for Schwarzschild spacetime, shaded regions denote the range of solutions for each spacetime, given by $\langle w \rangle_{i} \pm \sigma_{i}$, where $\langle w \rangle_{i}$ is the mean value and $\sigma_{i}$ is the 
standard deviation of the physical quantities. 
}
\label{fig:Density_all}
\end{figure*}
    
\begin{figure}
\centering
\includegraphics[width=0.475\textwidth]{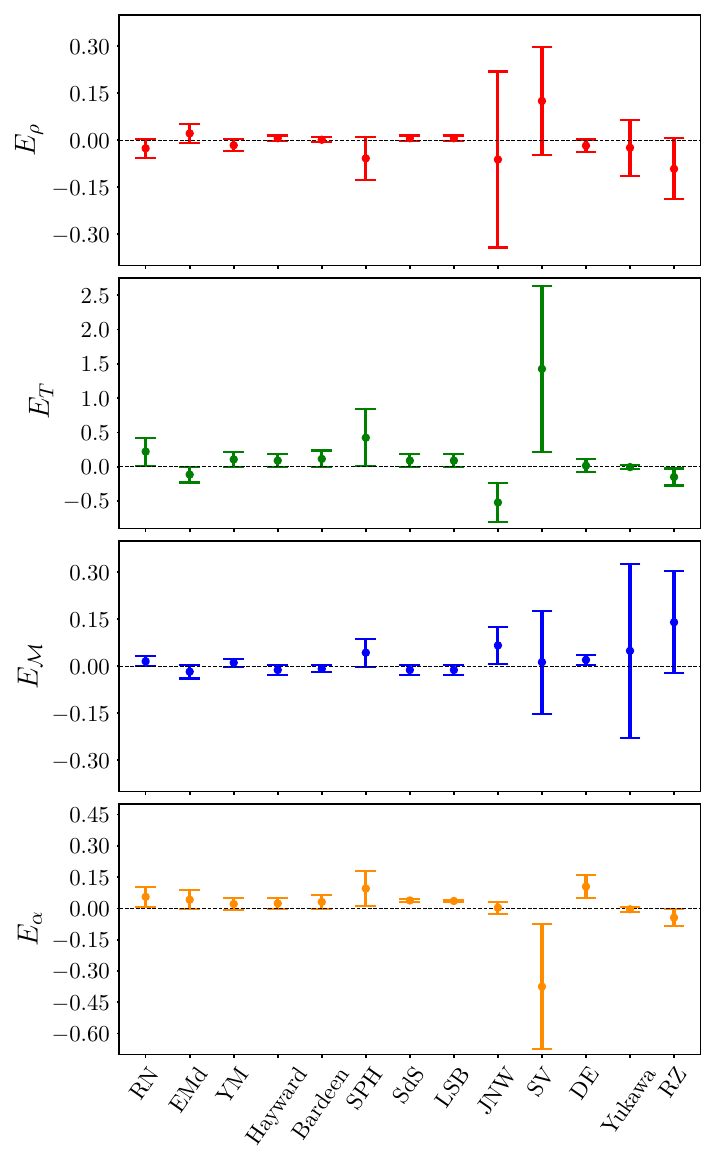}
\caption{Averaged relative error between spherical accretion in Schwarzschild spacetime and other black hole solutions,  defined as $E_{w_{i}}=\langle(w_{i}-w_{\rm Schw})/w_{\rm Schw}\rangle$, where $w_{i}=\rho,\, T,
\, {\cal M},\, \alpha$, correspond to the rest-mass density, temperature, Mach number, and the power-law index of the radial density profile.}
\label{fig:deviations}
\end{figure}
    
    The morphology of the spherically symmetric accretion flow onto thirteen non-rotating 
    black holes, obtained by solving equation \eqref{eq:generaleq}, is shown in Figure 
    \ref{fig:Density_all}, where the rest-mass density, temperature, radial velocity, and Mach 
    number defined as ${\cal M}\equiv V/c_{s}$, are plotted as a function of the normalized 
    radial coordinate, $r/r_{\rm EH}$. For convenience and comparison purposes, 
    we depict only the global behavior of the rest-mass density as the degrees of freedom of each 
    spacetime vary. The shaded regions are bounded by $\langle \rho \rangle_{i} \pm \sigma_{i}$, 
    where $\langle \rho \rangle_{i}$ is the mean rest-mass density and $\sigma_{i}$ is the 
    corresponding standard deviation, the Schwarzschild solution is shown with black line (complete set of solutions are presented in Appendix \ref{app:allsolutions}). 
     To quantify deviations of the Michel solution between Schwarzschild and thirteen black 
     hole spacetimes, we compute the averaged relative error, 
     $E_{w_{i}}=\langle(w_{i}-w_{\rm Schw})/w_{\rm Schw}\rangle$, for the plasma quantities 
     $w_{i}=\rho, T, {\cal M}, \alpha$, corresponding to the rest-mass density, temperature, 
     Mach number, and the power-law index of the radial density profile, respectively. 
     The average is computed over the radial coordinate, covering all solutions obtained by varying the free parameters of each spacetime. The results 
     for all spacetimes are shown in Figure \ref{fig:deviations}.
     The density 
     profile near the black hole shows the radial profile $\rho \sim r^{-\alpha}$ (for Schwarzschild we have $\alpha=1.327$). Among the spacetimes considered, the JNW and Yukawa solutions show the closest agreement with the Michel solution, with mean deviations of less than $1\%$. The RN, EMd, YM, Hayward, Bardeen, SdS, LSB, and RZ solutions exhibit mean deviations of approximately $2$--$5\%$, while the SPH and DEH solutions show mean deviations of approximately $10\%$. The SV wormhole exhibits a more pronounced deviation of approximately $37\%$. This behavior is associated with the traversable-wormhole nature of the SV solution, for which increasing the parameter $\tilde{\alpha}$ causes the density profile to become progressively flatter.

     On the other hand, far away from the black hole we observe exponencial differences for spacetimes with cosmological constant--such as SdS, 
    DEH and in those involving LSB. In this work we 
    have adopted relatively large values of $\Lambda$ to enhance the effects of the 
    cosmological constant on the accretion flow \citep[see for example][]{Mach_2015,
     Ahmed_2016}. However, to model correctly the actual universe, the estimated 
    value should be $\Lambda_{\rm cgs} \sim 10^{-56}\,\rm cm^{-2}$ 
    \citep{Schmidt_1998, Riess_1998, Perlmutter_1999} \footnote{The transformation 
    from geometric to cgs units can be done as following $\Lambda_{\rm cgs}=\Lambda_{\rm geo}(1.47651\times10^5M/M_\odot)^{-2}$.}.

    As the plasma accretes onto the black hole, we observe that the temperature--defined 
    as $T = p/\rho$ in code units and shown in the second column of the 
    figure--increases as the gas falls radially inward. In most cases, the temperature 
    rises with varying spacetime parameters and tends to be higher than in the 
    standard Schwarzschild spherical accretion scenario. In the SV 
    wormhole, the temperature increases uniformly even at large distances from 
    the black hole or throat, here we found largest differences with Schwarzschild case, of around  $\sim 150\%$ (see the second panel of Figure \ref{fig:deviations}).
    However, in the JNW spacetime, the gas remains cooler than in 
    Schwarzschild, with averaged relative error of $\sim 50\%$. For the RZ metric, we find that certain values of the 
    free parameter $a_{i}$ also lead to cooler gas compared to Schwarzschild. 
    
    As expected, the most significant deviations occur in spacetimes 
    that incorporate dark energy or a cosmological constant. In these cases--such 
    as the LSB, DEH, and SdS spacetimes--we observe an exponential increase in temperature with 
    radius, moving away from the black hole. Overall, our results show that the 
    temperature of the accreting flow can be significantly altered--either increased or 
    decreased--as a direct consequence of changes in the spacetime geometry.
    
    In the third panel of Figure \ref{fig:Density_all}, we present 
    the radial velocity profiles of the accreting flow. The overall behavior is similar 
    to that of the rest-mass density: in most cases, the velocity profiles are in good 
    agreement with the Michel accretion solution for a Schwarzschild black hole. 
    As expected, the gas accelerates as it falls toward the black hole, reaching 
    radial velocities of up to  $|V^r |\sim 0.4$. Regular and charged black holes, as 
    well as the RZ metric, exhibit only small deviations from the 
    Schwarzschild Michel solution. In contrast, spacetimes such as SPH, LSB, Yukawa-like, and SV  
    wormhole metrics exhibit higher infall velocities. In the SV wormhole 
    case, the gas shows larger velocities at large distances; however, as it approaches 
    the black hole, the velocity decreases, indicating a more complex and dynamic 
    accretion flow. On the other hand, we observe a significant suppression of radial 
    motion in the SdS and DEH spacetimes, where 
    the gas effectively freezes at large distances due to the effect of the cosmological 
    constant. A similar freezing behavior is observed near the pseuso-event horizon in the 
    JNW spacetime, where the radial velocity approaches zero.
    
    We compute the Newtonian Mach number (see the fourth panel of Figure \ref{fig:Density_all}) to compare our results with Michel accretion onto a Schwarzschild black hole. In the Michel solution, the flow transitions from subsonic to supersonic at the critical point and reaches $\mathcal{M} \sim 3.2$ near the event horizon. In general, the Mach number decreases with radius. The most 
    notable deviations from the Schwarzschild case occur at large radii in the 
    SdS, LSB, and DEH spacetimes, where the Mach number remains close to zero, indicating a nearly 
    stagnant flow. In Yukawa-like spacetimes, consistent with the behavior of the 
    radial velocity, the Mach number is higher than in the Schwarzschild case. 
    Near the black hole, we typically observe $\mathcal{M} \leq 3.2$, however, 
    in certain cases--such as SdS, LSB, RZ, and DEH spacetimes--the 
    Mach number exceeds this value, reaching $\mathcal{M} \geq 3.2$. 
    On the other hand, in the SV spacetime, the flow behaves similarly 
    to the Schwarzschild case; however, in the wormhole regime, the gas Mach 
    number remains close to unity, in contrast to the supersonic values seen in 
    the Michel solution for a Schwarzschild black hole.
    
    \section{Mass Accretion rates}\label{sec:accretion_rates}
    
    The gas morphology obtained in our models provides valuable insight into 
    how different spacetime geometries influence the accretion flow. Since the 
    flow is in a steady-state equilibrium around the black hole, it is important to 
    measure the mass accretion rate, as this offers a quantitative description of 
    how the black hole is fed by the surrounding gas. The mass accretion rate is calculated as follows:
    \begin{equation}
    \dot M=4\pi\rho U^r\sqrt{-g},
    \label{eq:mref}
    \end{equation}
    which is valid at any radial position, as it corresponds to a conserved quantity 
    derived from equation \eqref{constantsC1C2}. We normalize the mass accretion rates using the analytical expression derived by \citet{Petrich1989}, thereby isolating the effects of the spacetime geometry:
    \begin{equation}
    \dot{M}_{\rm c}=4\pi\lambda \rho_{\rm c} \sqrt{M}r_{acc}^{3/2},
    \label{eq:mdotref}
    \end{equation}
    where $\rho_{\rm c} $ is the rest-mass density at the critical point, and $\lambda$ 
    is a dimensionless coefficient that depends on the adiabatic index, taking values 
    of $\lambda \simeq 0.71$ for $\tilde{\gamma}=4/3$ and $\lambda \simeq 0.25$ for 
    $\tilde{\gamma}=5/3$. The accretion radius is defined as $r_{\rm acc}={M}/{(c_{s,\rm c}^2
    +V_{\rm c}^2)}$, where $c_{s,\rm c}$ and $V_{\rm c}$, are the sound speed and 
    radial three-velocity at the critical point $r_c$, respectively. Additionally, to enable a direct comparison of accretion rates across different solutions and 
    with the case of spherical accretion onto a Schwarzschild black hole, we reproduce 
    the Michel solution. The normalized mass accretion rate 
    for a Schwarzschild black hole, $\dot{M}_{S}$ is thus given by:
    \begin{equation}
    \dot{M}_{S} \equiv { \dot{M} }/{ \dot{M}_{\rm c}} = 1.276,
    \label{eq:MdotS}
    \end{equation}
    where $\dot{M}_{\rm c}$ corresponds to the reference accretion rate from equation 
    \eqref{eq:mdotref}. The normalized accretion rate for the fiducial spacetime given 
    in equation \eqref{eq:MdotS}, \ie Schwarzschild black hole will be used to assess how the mass 
    accretion rate increases or decreases for different families of black hole solutions.
    Since most of the spacetimes analyzed in this work reduces to Schwarzschild black 
    hole when the characteristic parameter vanishes, $\dot{M}_{S}$ serves as a limit value 
    of the mass accretion rate in those cases.
    
    The results for the mass accretion rates of all thirteen black hole solutions are 
    summarized in the first and third columns of Figure \ref{fig:Mdot_L_fits} in Appendix \ref{app:allsolutions}. 
    Using the Michel accretion solution for Schwarzschild black hole as a fiducial 
    model --indicated by the grey dashed line-- we identify three main accretion 
    scenarios: i) lower accretion rates than Schwarzschild for spacetimes with 
    charge (RN, EMd), a cosmological constant (SdS, LSB), JNW, and SPH; ii) intermediate accretion rates, 
    where the Schwarzschild case lies within the range of values, such as in the, 
    DEH, YM and RZ spacetimes; and iii) higher accretion rates 
    than Schwarzschild for regular black holes, SV wormhole and Yukawa-like spacetimes.
    The mass accretion rates asymptotically approach the Schwarzschild value in 
    all scenarios. We identify particularly interesting cases where accretion 
    is strongly suppressed, for example, in spacetimes with a large cosmological constant 
    (SdS) and LSB geometries. These scenarios may be relevant for supermassive 
    black holes in active galactic nuclei, where sub-Eddington accretion rates are 
    expected.
    
    \section{Bremsstrahlung Luminosity}\label{sec:bremsstrahlung}
    
    In this section, we analyze the bremsstrahlung emissivity produced by electron--proton collisions during the accretion process onto black holes. The bremsstrahlung emission provides a first step toward understanding how the spacetime geometry influences the resulting luminosity. We compute the bremsstrahlung luminosity from the corresponding equilibrium accretion solutions using the following expression \citep{Rybicki_Lightman1986,Zanotti2010,Cruz2020b}:
    
    \begin{equation}
    L_{\rm BR}=3\times10^{78}\int\sqrt{T}\rho^2\Gamma \sqrt{ \bar{\gamma}} 
    dV\left(\dfrac{M_\odot}{M}\right)\rm{erg\ s^{-1}}. \label{eq:Luminosity}
    \end{equation}
    
    Here, $\bar{\gamma}$ denotes the determinant of spatial 3-metric, and 
    $\Gamma$ is the Lorentz factor. The factor in front of the integral arises 
    the unit conversion\footnote{The luminosity is converted from geometrized 
    to physical (cgs) units as $L_{\rm cgs} = 3.62849 \times 10^{59} 
    \left(\frac{G}{c^{5}}\right) L_{\rm geo}\  {\rm erg / s}$, where $G$ is the 
    gravitational constant, and $c$ is the speed of light.}. 
    To enable a proper comparison we normalize the bremsstrahlung luminosity by the Eddington luminosity
    \footnote{The Eddington luminosity is defined as $L_{\rm Edd} = 1.26\times
    10^{38}\left(\dfrac{M}{M_\odot}\right)\rm{erg\ s^{-1}}$, where $M$ is the mass 
    of the black hole and $M_{\odot}$ is the solar mass.}, allowing our results 
    to be applicable across different black hole solutions and mass scales. We also 
    use the Schwarzschild black hole case as a fiducial model to quantify the effects 
    of spacetime geometry. 
For the luminosity analysis, we adopt the density and temperature corresponding 
to the scenario studied by \citet{Cruz2020b}, namely a stellar-mass black hole with 
$M_{\rm BH}=4M_\odot$ and a rest-mass density of $\rho_{\rm cgs}=
9.51\times 10^{-9}\, \rm g\,cm^{-3}$. 
The luminosity for the fiducial model (the Schwarzschild black hole) is:
    \begin{equation}
    L_{S} \equiv \log_{10}(L_{\rm BR}/L_{\rm Edd})=25.427.
    \label{eq:l0}
    \end{equation}
    In Figure \ref{fig:Mdot_L_fits} of Appendix \ref{app:allsolutions} (second and fourth columns), we present 
    the bremsstrahlung luminosity for all the spacetimes considered in this 
    study. The Schwarzschild case is shown as a grey dashed line or grey 
    plane, with its value given by equation \eqref{eq:l0}. We identify three main luminosity behaviors:
    i) Lower bremsstrahlung luminosity than Schwarzschild, observed in two of the
    spacetimes with charge (RN, EMd), regular black holes (Bardeen, Hayward), SPH, SV wormhole and Yukawa-like geometries; ii) Intermediate luminosity, where the Schwarzschild case lies within the overall range 
    of values, as seen in the LSB, YM and RZ metrics;
    and iii) Higher luminosity than Schwarzschild, found in DEH, SdS and in the JNW spacetime.\\

    The dependence of the mass accretion rates and luminosities on the spacetime parameters is captured by analytical fitting formulas. We computed 316 solutions for spacetimes characterized by a single free parameter and $\sim 4000$ solutions for those with two free parameters. The resulting accretion rates and luminosities were then fitted for each spacetime, yielding the fitting formulas reported in Table~\ref{tab:Mdot_L_fittings}.    
    
      \begin{table*}[htb]
    \centering
    \setlength{\tabcolsep}{0pt}
    \caption{Mass accretion rate and bremsstrahlung luminosity fitting formulas. We express the mass accretion rate in terms of $\dot{M}_{\rm c}$ given in equation \eqref{eq:mref}, and Schwarzschild mass accretion rate, $\dot M_{S}$. Similarly, the bremsstrahlung luminosity is normalized by the Eddington luminosity and luminosity on Schwarzschild black holes,  $L_{S}$, respectively. The fitting obtained from solutions showed in Figure \ref{fig:Mdot_L_fits}. For convenience, in Reisnner-Nordström spacetime we use $\tilde{Q}^2 \equiv Q^2+P^2$.} 
    \label{tab:Mdot_L_fittings}
    \hspace*{-0.5cm}
        \begin{tabular}{|l|l|l|}
        \hline\hline
        \hspace*{1mm}Black Hole Solution & \hspace*{1mm}$\dot{M}/\dot{M}_{\rm c} = $ & \hspace*{1mm}$\log_{10}(L_{\rm BR}/L_{\rm Edd})=$\\\hline\hline
    \hspace*{0.1mm} Reissner-Nordström & 
    \hspace*{1mm}$\dot M_{S} -0.1697\tilde Q^2 -0.0571\tilde Q^4$ & 
    \hspace*{1mm}$L_{S} -0.4090\tilde Q^2 + 0.1599\tilde Q^4 -0.3907\tilde Q^6$
     \\\hline
    \hspace*{0.1mm} Einstein-Maxwell dilaton & \mytable{l}{
    $\dot M_{S}+ 0.086P -0.014Q -1.611P^2-$ \\
    $0.020Q^2 +0.069PQ -0.277P^2Q^2$} & 
    \mytable{l}{
    $L_{S} + 0.079P + 0.007Q -2.654P^2 -0.071Q^2-$ \\
    $0.124PQ + 1.583P^2Q^2 -3.717P^3Q^2 -2.292P^3Q^3$
    }\\\hline
    \hspace*{0.1mm} Yang Mills & \mytable{l}
    {$\dot{M}_{S} -1.82Q_{\rm YM} -0.328Q^2 + 3.322Q_{\rm YM}^2 + $ \\
    $0.139Q^4 + 0.082Q_{\rm YM}Q^2 -4.24 Q_{\rm YM}^3$} & \mytable{l}{$L_{S} -1.644Q_{\rm YM} -0.331Q^2 + 0.860Q_{\rm YM}^2-$ \\
$0.220Q^4 -1.020Q_{\rm YM}Q^4 -0.423Q_{\rm YM}^2Q^4$} \\\hline
    \hspace*{0.1mm} Hayward & \hspace*{1mm}$\dot{M}_{S} -0.012\ell+ 0.209\ell^2 -0.477\ell^3+ 0.540\ell^4$ & \hspace*{1mm}$L_{S} + 0.0029\ell -0.3388\ell^2 + 0.6053\ell^3 -2.3018\ell^4$\\\hline
    \hspace*{0.1mm} Bardeen & \hspace*{1mm}$\dot{M}_{S} + 0.0149P^2 -0.0272P^4 + 0.1435P^6$ & \hspace*{1mm}$L_{S} -0.4535P^2 -0.0936P^4 -0.6520P^6$
    \\\hline
   \hspace*{0.1mm} Super-Planckian hair & \hspace*{1mm}$\dot{M}_{S} -0.908\ell_{\rm S} -5.719\ell_{\rm S}^2 + 34.425\ell_{\rm S}^3 -88.308\ell_{\rm S}^4$\hspace*{1mm}
     & \hspace*{1mm}$L_{S} -2.1680\ell_{\rm S} -1.6652\ell_{\rm S}^2 -5.0825\ell_{\rm S}^3 -22.9798\ell_{\rm S}^4$ \\\hline
    \hspace*{1mm}Schwarzschild-de Sitter & \hspace*{1mm}$\dot{M}_{S} -738.26\Lambda +75270.49\Lambda^2$ & \hspace*{1mm}$L_{S} + 12790.48\Lambda + 67040132.85\Lambda^2$ \\\hline
    \hspace*{1mm}Lorentz symmetry breaking\hspace*{1mm} & \mytable{l}
    {$\dot M_{S} -1036.87\Lambda -1.49\lambda_{\rm LSB} +$\\ $3905243.59\Lambda^2\lambda_{\rm LSB}+2726\Lambda\lambda_{\rm LSB}^2-$\\
    $8311907.39\Lambda^2\lambda_{\rm LSB}^2 +\Lambda^3\lambda_{\rm LSB}^2$} & 
    \mytable{l}
    {
    $L_{S} + 25922.518\Lambda -3.369\lambda_{\rm LSB}+ 7.596\lambda_{\rm LSB}^2 -$\\
    $9.315\lambda_{\rm LSB}^3 -64006.829\Lambda\lambda_{\rm LSB} -4063074.472\Lambda\lambda_{\rm LSB}^2 -$\\
    $10188.974\Lambda\lambda_{\rm LSB}^4 + 3353289.882\Lambda^2\lambda_{\rm LSB} +$\\
    $ 4109798.651\Lambda\lambda_{\rm LSB}^2+ \Lambda^3\lambda_{\rm LSB}+ \Lambda^4\lambda_{\rm LSB}$}\\\hline
    \hspace*{1mm}Janis-Newman-Winicour & \hspace*{1mm}$\dot M_{S} -0.288 + 0.244\gamma + 0.044\gamma^2$ & \mytable{l}{$L_{S} + 5.84 -60.499\gamma + 457.45\gamma^2 -2116.29\gamma^3 +$\\
    $5869.89\gamma^4 -9831.71\gamma^5 + 9726.21\gamma^6 -5224.35\gamma^7 +$\\ $1173.46\gamma^8$\\
    } \\\hline
    \hspace*{1mm}Simpson-Visser Wormhole & \hspace*{1mm}$\dot M_{S} + 0.086\tilde a -0.030\tilde a^2+ 0.030\tilde a^3$ & 
    \hspace*{1mm}$L_{S}-0.177\tilde a -0.073\tilde a^2+0.020\tilde a^3$ \\\hline
    \hspace*{1mm}Dark energy Hayward & \mytable{l}
    {$\dot{M}_{S} +\alpha\omega -0.0093\ell + 0.1785\ell^2 -0.3909\ell^3 +$\\
    $0.4665\ell^4$} & 
    \mytable{l}
    {
    $10^{(\log_{10}L_{S}+\alpha\omega-0.0017\ell+0.0011\ell^2 -0.0042\ell^3-0.0070\ell^4 )}$
    }\\\hline
    \hspace*{1mm}Yukawa & \mytable{l}{$\dot M_{S}+1.716\exp\{-(0.0005(\lambda-40)^2+$\\
    $32(\delta+0.85)^2)\}$} & 
    \mytable{l}{$L_{S}-0.28\exp\{-(0.0002(\lambda-40)^2+10(\delta+0.85)^2)\}$}\\\hline
    \hspace*{1mm}Rezzolla-Zhidenko & \hspace*{1mm}$\dot M_{S} -0.6909a_0 -0.2192a_1 + 0.0665a_0a_1 $
     & \mytable{l}
    {
    $L_{S} -0.2669a_0 -0.1188a_1 +0.0068a_0a_1-$ \\
    $0.0877a_0^2a_1 +0.0009a_0a_1^2 -0.0281a_0^2a_1^2$
    } \\\hline
    \end{tabular}
    \end{table*}    

    \section{Astrophysical Implications}\label{sec:astro}
  
Accretion onto black holes represents the most efficient astrophysical mechanism for energy extraction. The infall of hot, magnetized plasma can produce intense electromagnetic emission, particularly at X-ray wavelengths, while the accreted matter contributes to the growth of the black hole.
The origin of supermassive black holes may involve a variety of seed populations, including primordial, stellar remnant, intermediate-mass, and massive black hole seeds formed in the early universe, followed by growth through accretion and mergers \citep{Lora-Clavijo2013,Fan2023,Bogdan2024,Maiolino2024,Huang2026}.\\
Recently, the LIGO-Virgo-KAGRA (LVK) gravitational-wave detectors detected the first binary black hole with component masses in the intermediate-mass regime, namely, GW190521, and subsequent observing campaigns have identified additional sources with even larger masses \cite{Abbott2020c,Abac2026}. The formation mechanism of these massive black holes remains an open question, with accretion being one possible evolutionary channel. In this section, we explore two scenarios: accretion rates onto supermassive black holes and accretion onto the progenitors of intermediate-mass black holes.

\subsection{Accretion rates onto supermassive black holes}
\label{sec:astro}  

Here, we estimate the mass accretion rates for supermassive black holes with masses in the range $M \in [10^{6},5\times10^{10}]\,M_{\odot}$, with the aim of comparing the accretion rates obtained for the thirteen solutions with those of the Schwarzschild spacetime. For these calculations, we use the fitting formulas derived in the previous section and summarized in Table~\ref{tab:Mdot_L_fittings}.

\begin{figure}
\centering
\includegraphics[width=0.49\textwidth]{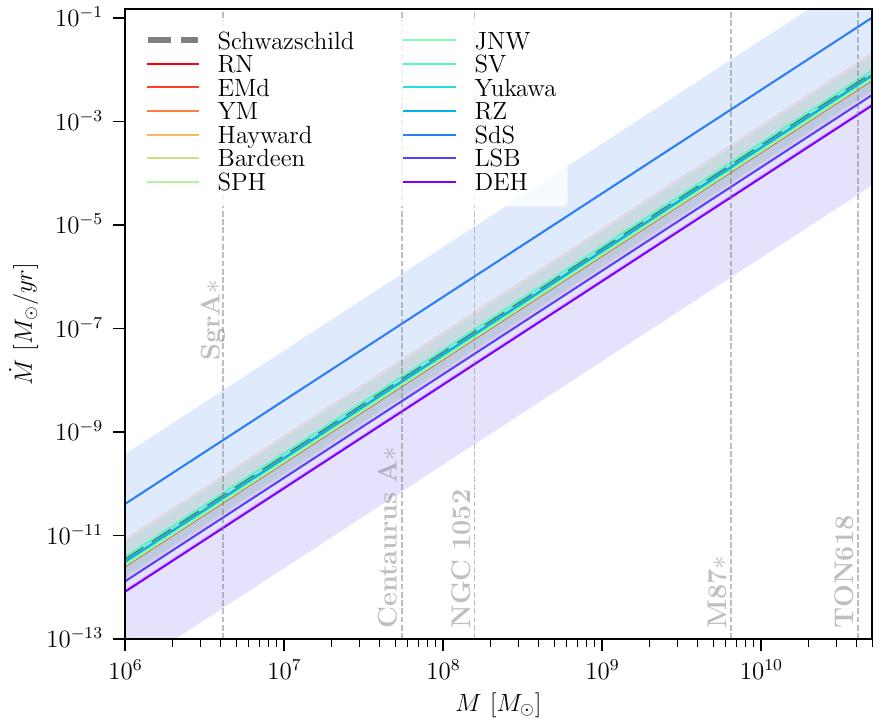}
\caption{Mass accretion rate as a function of black hole mass, focusing on 
supermassive black holes with masses in the range $M \in [10^{6},5\times10^{10}]
\,M_{\odot}$. The lines show the mean values, while the shaded regions 
indicate the standard deviations obtained by varying the parameters of 
each black hole solution. The Schwarzschild case is shown as a gray 
dashed line. Vertical dashed lines indicate representative 
black hole masses inferred from observations.}
\label{fig:smbh}
\end{figure}

In Figure~\ref{fig:smbh}, we present the mass accretion rates, in units of solar masses per year, for all black hole solutions. The Schwarzschild case is represented by the dashed grey line. For each spacetime and a given black hole mass, we calculate the accretion rates while varying the corresponding spacetime parameters (see Table~\ref{tab:spaceparameters}). We then compute the mean accretion rate, represented by the solid lines, together with the corresponding standard deviation, shown by the shaded regions. We also include several representative sources, namely the supermassive black holes Sgr~A* ($M=4.14\times 10^{6} \,M_{\odot}$ \cite{EHT_SgrA_PaperI_etal}), Centaurus A* ($M=5.5\times 10^{8} \,M_{\odot}$ \cite{Janssen2021}), NGC~1052 ($M=1.585\times 10^{8} \,M_{\odot}$ \cite{Baczko2024}), M87* ($M=6.5\times 10^{9} \,M_{\odot}$ \cite{EHT_M87_PaperI}), and TON~618 ($M=4.07\times 10^{10} \,M_{\odot}$ \cite{Xue2019}). In particular, we focus on Sgr~A* and M87*, for which the Event Horizon Telescope (EHT) Collaboration estimates mass accretion rates of $(5.2-9.5) \times 10^{-9}\,M_{\odot}\,\mathrm{yr}^{-1}$ and $(3-20) \times 10^{-4}\,M_{\odot}\,\mathrm{yr}^{-1}$, respectively. 

Most of the models exhibit only moderate deviations from the Schwarzschild case. However, the SdS and LSB spacetimes show substantially larger differences, with deviations of up to approximately two orders of magnitude in the mass accretion rates. Overall, the 13 spacetimes can be grouped according to the magnitude of their deviations from the Schwarzschild accretion rate. The SdS, LSB, and DEH spacetimes exhibit deviations of $\gtrsim 50\%$, while the SPH, SV, YM, and RZ spacetimes show intermediate deviations of $\gtrsim 5\%$ and $\lesssim 50\%$. In contrast, the RN, EMd, Hayward, Bardeen, JNW, and Yukawa spacetimes produce deviations of $\lesssim 5\%$ relative to the Schwarzschild case. These results indicate that, for most of the considered spacetimes, the accretion rates onto supermassive black holes remain relatively close to the Schwarzschild prediction, whereas the SdS, LSB, and DEH metrics can lead to significant modifications. Given that the current uncertainties in EHT measurements are of order $17\%$ \cite{Kocherlakota2021b}, deviations of this magnitude could have a significant impact on the inferred accretion rates and may be potentially detectable with sufficiently precise observations.

\subsection{Progenitors of intermediate-mass black holes}\label{sec:astro}
\begin{figure}
\centering
\includegraphics[width=0.465\textwidth]{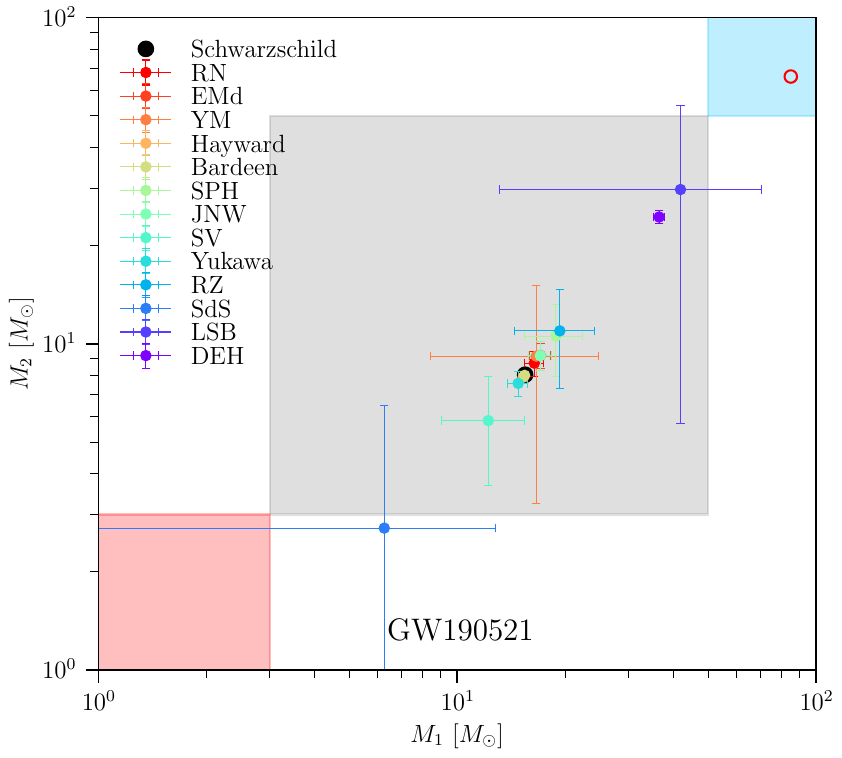}
\caption{Mass--mass diagram of the progenitor black holes for the binary 
merger GW190521. The red circle denotes the component masses inferred 
by LIGO--Virgo. Black dot correspond to progenitor masses obtained 
assuming a Schwarzschild black hole, while colored dots show the results 
for the different black hole spacetimes. Error bars indicate the variations 
produced by changing the parameters of each spacetime. The shaded red, 
gray, and blue regions denote the mass ranges of primordial black holes 
($M \leq 3\,M_{\odot}$), stellar-mass black holes ($3\,M_{\odot} < M < 50
\,M_{\odot}$), and intermediate-mass black holes ($50\,M_{\odot} < M < 
150\,M_{\odot}$), respectively.}
\label{fig:imbh}
\end{figure}

The second astrophysical scenario we consider is the estimation of the progenitor masses of the binary black hole merger event GW190521, as reported by the LVK Collaboration \citep{Abbott2020c}. For this purpose, we assume that each black hole grows through accretion from its formation until the time of merger. The mass accretion rates of the two components are given by:
\begin{eqnarray}
\dot{M}_{1}=\frac{\dot{M}_{\rm bin}}{\sqrt{2(1+q)}}\,, \qquad \dot{M}_{2}=\frac{\dot{M}_{\rm bin} \sqrt{q}}{\sqrt{2(1+q)}}\,, \label{eq:binaryacc}
\end{eqnarray}
where $q=M_{2}/M_{1}\leqslant 1$ is the binary mass ratio, and $\dot{M}_{\rm bin}$ is the mass accretion rate of the binary system. In deriving these expressions, we neglect contributions from the orbital eccentricity and assume that the binary orbital period is much shorter than the accretion timescale. We compute $\dot{M}_{\rm bin}$ for each spacetime using the fitting formulas summarized in Table~\ref{tab:Mdot_L_fittings}.

The binary accretion rate $\dot{M}_{\rm bin}$ depends on the characteristic accretion rate $\dot{M}_{\rm c}$, defined in Equation~\eqref{eq:mdotref}. We adopt the redshift dependent rest-mass density and sound speed of the intergalactic medium given by
\begin{eqnarray}
\rho_{c}(z)=& 200 m_{\rm H}  \left(\dfrac{1+z}{10^{3}}\right)^{3} \rm\dfrac{g}{ cm^3}\,,  \label{eq:den}\\
c_{s,c}(z)=& 5.7 \sqrt{\dfrac{1+z}{10^{3}}} \left[1 + \left(\dfrac{1+z_{\rm dec}}{1+z}\right)^{\beta} \right]^{-\frac{1}{2\beta}} \rm \dfrac{km}{s}, \label{eq:cs} \\ 
V_{c}(z)=& 17\sqrt{\dfrac{1+z}{10}}\left( \dfrac{M_{2\sigma}}{10^{8} M_{\odot}} \right)^{1/3}  \rm\dfrac{km}{s},
\end{eqnarray}
where $M_{2\sigma}=10^{17} e^{-5.57(1+z)^{0.57}} M_{\odot}$, $\beta=1.72$, $z_{\rm dec}=130$ is the redshift at which the baryonic matter decouples from the radiation fluid, and $m_{\rm H}=1.6735575\times10^{-24} \mathrm{g}$ is the hydrogen mass. The sound speed prescription assumes adiabatic cooling of the plasma after $z=100$, when the baryonic gas thermally decouples from the cosmic microwave background (CMB). For further details, see \cite{Ricotti2008,DeLuca2020,Cruz2021a}.

Following \citet{Cruz2021a}, we perform a backward integration from the redshift of the binary merger, $z=0.82$, as estimated by the LVK Collaboration, to $z=100$. We integrate Equation~\eqref{eq:binaryacc} for all 13 black hole spacetimes, varying the corresponding spacetime parameters listed in Table~\ref{tab:spaceparameters}. We use a redshift resolution of $dz=10^{-3}$ and adopt the component masses at $z=0.82$ as $M_{1}=85^{+21}_{-14}\,M_{\odot}$ and $M_{2}=66^{+17}_{-18}\,M_{\odot}$. 

In Figure~\ref{fig:imbh}, we show the inferred progenitor masses of the two components of GW190521 at $z=100$. In the mass--mass diagram, the red, gray, and blue shaded regions represent three characteristic black hole mass ranges: primordial black holes, $M \leq 3\,M_{\odot}$; stellar mass black holes, $3\,M_{\odot}<M \leq 50\,M_{\odot}$; and intermediate mass black holes, $M>50\,M_{\odot}$, respectively. The red open circle indicate the component masses at the merger redshift, $z=0.82$. The colored points represent the mean progenitor masses obtained for each spacetime, with the corresponding standard deviations arising from the variation of the spacetime parameters. The Schwarzschild case is represented by the black point, yielding progenitor masses of $M_{1}=15.46\,M_{\odot}$ and $M_{2}=8.05\,M_{\odot}$.

The progenitor masses obtained for the Bardeen and Hayward spacetimes are very similar to those of the Schwarzschild case, with deviations of less than $1\%$ and $2.5\%$, respectively. Smaller progenitor masses are obtained for the Yukawa, SV, and SdS spacetimes. The largest decrease is found for the SdS spacetime, for which our calculations suggest that the secondary component could lie in the primordial black hole mass range, with mean progenitor masses of $M_{1}=6.26\,M_{\odot}$ and $M_{2}=2.73\,M_{\odot}$. 
On the other hand, several spacetimes yield larger progenitor masses than the Schwarzschild case, namely RN, EMd, JNW, SPH, RZ, YM, DEH, and LSB. The largest deviations are obtained for the DEH and LSB spacetimes, for which the inferred progenitor masses reach $M_{1}>36\,M_{\odot}$ and $M_{2}>29\,M_{\odot}$, respectively. These results illustrate that the assumed spacetime geometry can significantly affect the inferred progenitor masses when the black hole growth history is reconstructed through the relativistic accretion.

    \section{Summary}\label{sec:conclusions}

    In this work, we have systematically investigated stationary, spherically symmetric accretion onto thirteen non-rotating black hole spacetimes spanning a broad range of theoretical scenarios, including charged and regular black holes, scalar and dilaton fields, dark energy, cosmological constants, modified gravity, and parametric deviations from Schwarzschild spacetime. We quantified the mass accretion rate and electron--proton bremsstrahlung luminosity for each model and derived analytical fitting formulas as functions of the corresponding spacetime parameters.
    
    Relative to the Schwarzschild solution, the radial equilibrium profiles of the hydrodynamic quantities generally exhibit deviations below $10\%$ for most models. Largest deviations are found for the SPH, JNW, SV, Yukawa, and RZ spacetimes, showing  that even modest modifications of the spacetime geometry can produce appreciable changes in the accretion flow. For supermassive black holes with masses in the range $M\in[10^{6},5\times10^{10}]\,M_{\odot}$, the accretion rates in the SdS, LSB and  DEH spacetimes differ by $\gtrsim20\%$ from the Schwarzschild case.
    
    We further applied the fitting formulas to the progenitor masses of the GW190521 event at $z=100$. The inferred masses show significantly larger deviations from the Schwarzschild prediction. In particular, the SV and SdS models yield progenitor masses approximately $20\%$ and $60\%$ smaller than the corresponding Schwarzschild predictions, $M_{1}=15.46\,M_{\odot}$ and $M_{2}=8.05\,M_{\odot}$, respectively. In contrast, DEH and LSB spacetimes produce progenitor masses $130\%$ larger than those obtained for Schwarzschild black hole.

 These results demonstrate that modifications of the spacetime geometry can leave measurable imprints on both the accretion dynamics and the associated bremsstrahlung luminosity. Although the deviations in local fluid quantities are generally modest, their cumulative effect on the accretion rate and inferred black hole masses can be substantial. Spherical accretion therefore provides a useful theoretical framework for probing potential signatures of deviations from the Schwarzschild geometry and, more broadly, for exploring observational consequences of alternative gravitational scenarios. Future work will extend this analysis to rotating black holes and incorporate more realistic plasma physics, including realistic equations of state, radiation, and viscosity.
       
    \section*{Acknowledgements}
    
This research was supported by Dirección General de Asuntos del Personal Académico, 
Universidad Nacional Autónoma de México grant IA103725 and by Ciencia B\'asica y 
de Frontera 2023-2024 program of SECIHTI M\'exico projects CBF2023-2024-1102, 
257435 and 1147615. 
Simulations were performed at Atocatl cluster of LAMOD-DGTIC-UNAM,  at Laboratorio 
Nacional de Supercómputo del Sureste de México project 202403062N, Miztli 
supercomputing cluster at Dirección General de Cómputo y Tecnologías de Información 
UNAM project LANCAD-UNAM-DGTIC-479, and local workstations 
“Going Merry" and “Thousand Sunny''.    
{

}
\begin{sidewaystable*}[h!]
\vspace{7cm}
\renewcommand{\arraystretch}{0.1} \setlength{\tabcolsep}{0pt}
\caption{Spacetime $g_{tt}$ and $g_{rr}$ metric components. We list the spacetime models in the first column, the $g_{tt}$ and $g_{rr}$ component in second and third, respectively. While the spacetime individual free parameters are reported in the fourth column. In Einstein-Maxwell dilaton, the terms $r_{\pm}$ are defined as $r_{\pm}=M\pm r_0$, $r_0=\displaystyle\sqrt{M^2+\Sigma^2-P^2-Q^2}$, where $\Sigma=(P^2-Q^2)/2M$.}
{\fontsize{9.0}{15}\selectfont
\begin{tabular}{|l|c|c|l|}
\hline\hline
Black Hole Solutions & $g_{tt}=$ & $g_{rr}=$ & Parameters\\\hline\hline
Schwarzschild & \mytable{l}{$-\left(1-\dfrac{2M}{r}\right)$} & \mytable{l}{$\left(1-\dfrac{2M}{r}\right)^{-1}$} & \mytable{l}{$M$ Mass} \\\hline
Reissner-Nordström \citep{Reissner1916,Nordstrom1918,Carroll2019} & \mytable{c}{$-\left(1-\dfrac{2M}{r}+\dfrac{Q^2+P^2}{r^2}\right)$} & \mytable{c}{$\left(1-\dfrac{2M}{r}+\dfrac{Q^2+P^2}{r^2}\right)^{-1}$} & \mytable{l}{$Q$ Electric charge \\ $P$ Magnetic charge} \\\hline
Einstein-Maxwell dilaton \citep{Kallosh1992,Hirschmann2018} & \mytable{c}{$-\dfrac{(r-r_+)(r-r_-)}{r^2-\Sigma^2}$} & \mytable{c}{$\left[\dfrac{(r-r_+)(r-r_-)}{r^2-\Sigma^2}\right]^{-1}$} & \mytable{l}{$\Sigma$ Dilaton charge} \\\hline
Yang-Mills \citep{RinconGomez2024} & \mytable{c}{$-\left(1-\dfrac{2M}{r}+\dfrac{Q^2}{r^2}+Q_{\rm YM}\right)$} & \mytable{c}{$\left(1-\dfrac{2M}{r}+\dfrac{Q^2}{r^2}+Q_{\rm YM}\right)^{-1}$} & \mytable{l}{$Q$ Electric charge \\ $Q_{\rm YM}$ Yang-Mills charge} \\\hline
Hayward \citep{Hayward2006,Malafarina2023} & \mytable{c}{$-\left(1-\dfrac{2Mr^2}{r^3+2\ell^2M}\right)$} & \mytable{c}{$\left(1-\dfrac{2Mr^2}{r^3+2\ell^2M}\right)^{-1}$} & \mytable{l}{$\ell$ Length scale} \\\hline
Bardeen \citep{ayonbeato2000,Bambi2014,Malafarina2023} & \mytable{c}{$-\left(1-\dfrac{2Mr^2}{(r^2+P^2)^{3/2}}\right)$} & \mytable{c}{$\left(1-\dfrac{2Mr^2}{(r^2+P^2)^{3/2}}\right)^{-1}$} & \mytable{l}{$P$ Magnetic charge} \\\hline
Super-Planckian hair \citep{Cadoni2023,Malafarina2023} & \mytable{c}{$-\left(1-\dfrac{2Mr^2}{(r+\ell_{\rm S})^3}\right)$} & \mytable{c}{$\left(1-\dfrac{2Mr^2}{(r+\ell_{\rm S})^3}\right)^{-1}$} & \mytable{l}{$\ell_{\rm S}$ Length scale} \\\hline
Schwarzschild-de Sitter \citep{deSitter_1917,Kottler1918,Carroll2019} & \mytable{c}{$-\left(1-\dfrac{2M}{r}-\dfrac{\Lambda r^2}{3}\right)$} & \mytable{c}{$\left(1-\dfrac{2M}{r}-\dfrac{\Lambda r^2}{3}\right)^{-1}$} & \mytable{l}{$\Lambda$ Cosmological constant} \\\hline
Lorentz symmetry breaking \citep{Yang2023,Araujo2024} & \mytable{c}{$-\left(\dfrac{1}{1-\lambda_{\rm LSB}}-\dfrac{2M}{r}-\dfrac{\Lambda r^2}{3(1-\lambda_{\rm LSB})}\right)$} & \mytable{c}{$\left(\dfrac{1}{1-\lambda_{\rm LSB}}-\dfrac{2M}{r}-\dfrac{\Lambda r^2}{3(1-\lambda_{\rm LSB})}\right)^{-1}$} & \mytable{l}{$\lambda_{\rm LSB}$ Lorentz-violation parameter \\ $\Lambda$ Cosmological constant} \\\hline
Janis-Newman-Winicour \citep{JanisNewmanWinicour1968,Virbhadra_1997} & \mytable{c}{$-\left(1-\dfrac{2M}{\gamma r}\right)^\gamma$} & \mytable{c}{$\left(1-\dfrac{2M}{\gamma r}\right)^{-\gamma}$} & \mytable{l}{$\gamma$ Scalar field intensity} \\\hline
Simpson-Visser Wormhole \citep{Simpson_2019} & \mytable{c}{$-\left(1-\dfrac{2M}{\sqrt{r^2+\tilde a^2}}\right)$} & \mytable{c}{$\left(1-\dfrac{2M}{\sqrt{r^2+\tilde a^2}}\right)^{-1}$} & \mytable{l}{$\tilde a$ Regularizer parameter} \\\hline
Dark energy-Hayward \citep{LiZhangHuang_2024} & \mytable{c}{$-\left(1-\dfrac{2Mr^2}{r^3+2\ell^2M}+\dfrac{\alpha}{r^{3\omega+1}}\right)$} & \mytable{c}{$\left(1-\dfrac{2Mr^2}{r^3+2\ell^2M}+\dfrac{\alpha}{r^{3\omega+1}}\right)^{-1}$} & \mytable{l}{$\ell$ Length scale \\ $\omega$ Dark energy parameter \\ $\alpha$ Normalization factor} \\\hline
Yukawa-like \citep{Ho1995,DeLaurentis2018,CruzOsorio2021} & \mytable{c}{$-\left(1-\dfrac{2M}{r}\dfrac{(\delta e^{-r/\lambda}+1)}{\delta+1}\right)$} & \mytable{c}{$1-\dfrac{2M}{r}\left[\dfrac{\delta e^{-r/\lambda}+1}{\delta+1}+\dfrac{\delta r e^{-r/\lambda}/\lambda-2}{\delta+1}\right]$} & \mytable{l}{$\lambda$ Scale parameter \\ $\delta$ Deviation parameter} \\\hline
Rezzolla-Zidenkho \citep{Rezzolla2014} & \mytable{c}{$-\left(1-\dfrac{2M}{r}\right)\left[1+a_0\left(\dfrac{2M}{r}\right)^2+a_1\left(\dfrac{2M}{r}\right)^3\right]$} & \mytable{c}{$\left(1-\dfrac{2M}{r}\right)^{-1}\left[1+a_0\left(\dfrac{2M}{r}\right)^2+a_1\left(\dfrac{2M}{r}\right)^3\right]^{-1}$} & \mytable{l}{$a_0,\ a_1$ Fitting parameters} \\\hline
\end{tabular}
}
\label{tab:spacetimes}
\end{sidewaystable*}

    \begin{table*} 
    \centering
    \setlength{\tabcolsep}{0pt}
    {\fontsize{7.6}{10}\selectfont
    \caption{Spacetime parameters range. For each spacetime, we show the parameters range to have the physical solution; black holes, naked singularities or wormholes. Additionally, we report the value for which we recover Schwarzschild spacetime. 
    } 
    \begin{tabular}{|l|c|c|c|}
    \hline\hline
    Black Hole Solution & Parameters & Limit case (Schwarzschild) & Naked singularity \\\hline\hline 
    Reissner-Nordström & $\tilde Q^2\in[0,0.98]$ & $\tilde Q=0$ & $\tilde Q^2\geq1$\\\hline
    Einstein-Maxwell dilaton & $P\in[0,0.4],\ Q\in[0,1]$ & $P=0$, $Q=0$ & $P^2+Q^2\gtrsim1$ \\\hline
    Yang-Mills & $Q_{\rm YM}\in[-0.5,0.5],\ Q^2\in[0,0.64]$ & $Q_{\rm YM}=0,\,Q=0$ & $1/(1+Q_{\rm YM})< Q^2$  \\\hline
    Hayward & $\ell\in[0,0.7]$ & $\ell=0$ & No naked singularity \\\hline
    Bardeen  & $P\in[0,0.68]$ & $P=0$ & No naked singularity \\\hline
    Super-Planckian hair  & $\ell_{\rm S}\in[0,0.29]$ & $\ell_{\rm S}=0$ & No naked singularity \\\hline
    Schwarzschild-de Sitter & $\log_{10}\Lambda\in[-10,-3.4]$ & $\Lambda=0$ & $\Lambda\gtrsim0.03$ \\\hline
    Lorentz symmetry breaking\hspace{1mm} & $\ \log_{10}\Lambda\in[-10,-3.4],\ \lambda_{\rm LSB}\in[0,0.96]\ $ & $\Lambda=0$, $\lambda_{\rm LSB}=0$ & $\lambda_{\rm LSB}>1$  \\\hline
    Janis-Newman-Winicour & $\gamma\in[0.04,1]$ & $\gamma=1$ & $\gamma\neq1$\\\hline
    Simpson-Visser Wormhole & $\tilde a\in[0,4]$ & $\tilde a=0$ & No naked singularity \\\hline
    Dark energy Hayward & $-1<\omega\leq-2/3,\ \ell\in[0,0.68]$ & $\ell=0,\ \alpha=0$ & No naked singularity \\\hline
    Yukawa-like & $\lambda\in[40,250],\ \delta\in[-0.85,0.85]$ & $\delta=0,\ r/r_{\rm EH}\gg1$ & No naked singularity  \\\hline
    Rezzolla-Zhidenko & $a_0\in[-0.2,0.7],\ a_1\in[-0.3,1]$ & $a_0=0,\ a_1=0$ & No naked singulariry \\\hline
    \end{tabular} \label{tab:spaceparameters} 
    }  
    \end{table*}

    \appendix
    \section{Families of spacetimes and parameters}
    \label{app:spacetimes}
      
     Table \ref{tab:spacetimes} presents the complete list of spacetimes 
    analyzed in this work. These are categorized as described in Section 
    \ref{sec:spacetimemodels}. 
    Table \ref{tab:spaceparameters} lists the parameter 
    ranges explored for each spacetime. For black hole solutions, we restrict 
    our analysis to parameter values that ensure the presence of an event 
    horizon, keeping the singularity hidden and allowing the use of a normalized 
    radial coordinate.
    \section{Radial profiles of physical quantities, accretion rates and luminosities.}
    \label{app:allsolutions}
\begin{figure*}
    \centering
    \includegraphics[width=0.494\textwidth]{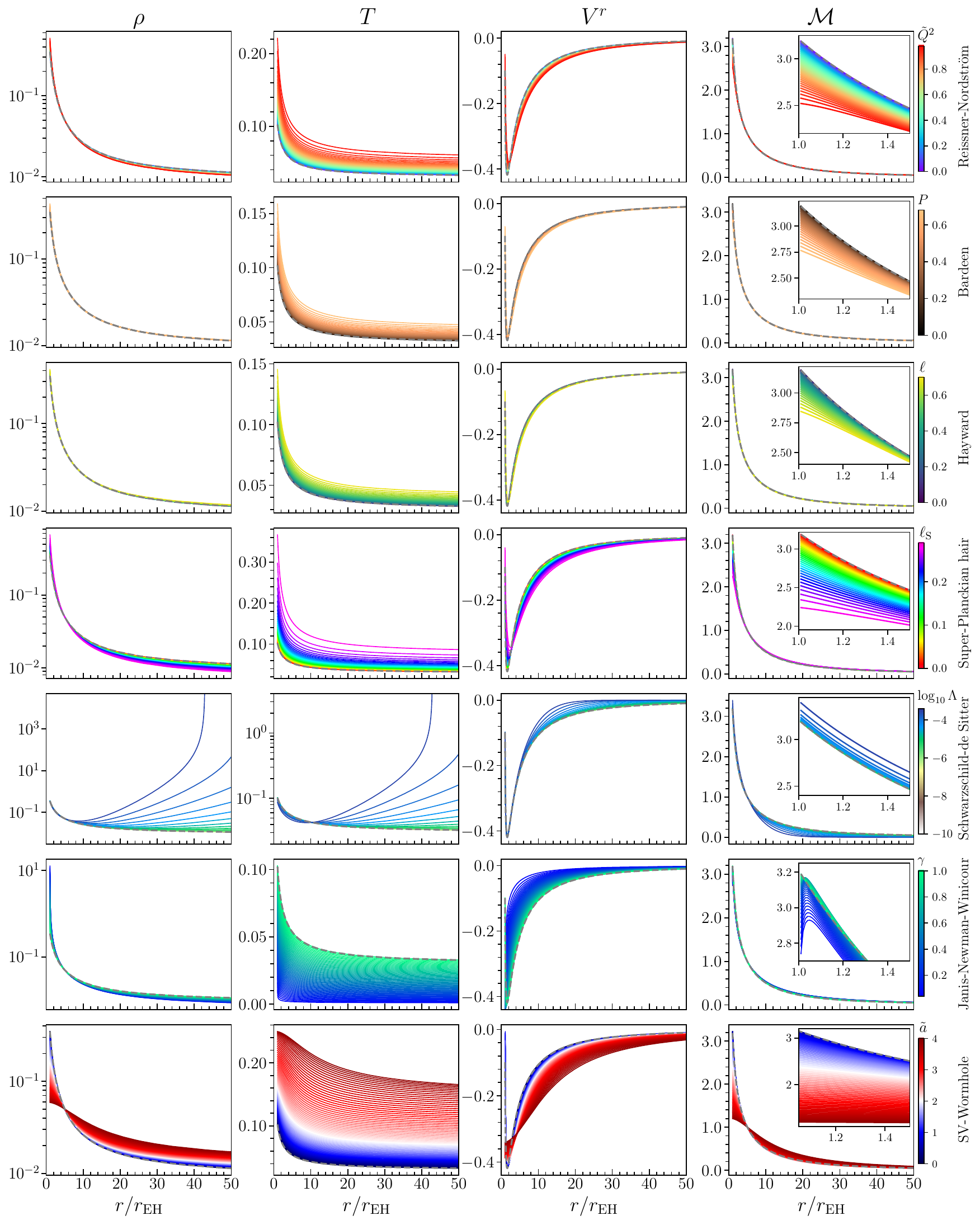}\hspace{-0.17cm}
    \includegraphics[width=0.502\textwidth]{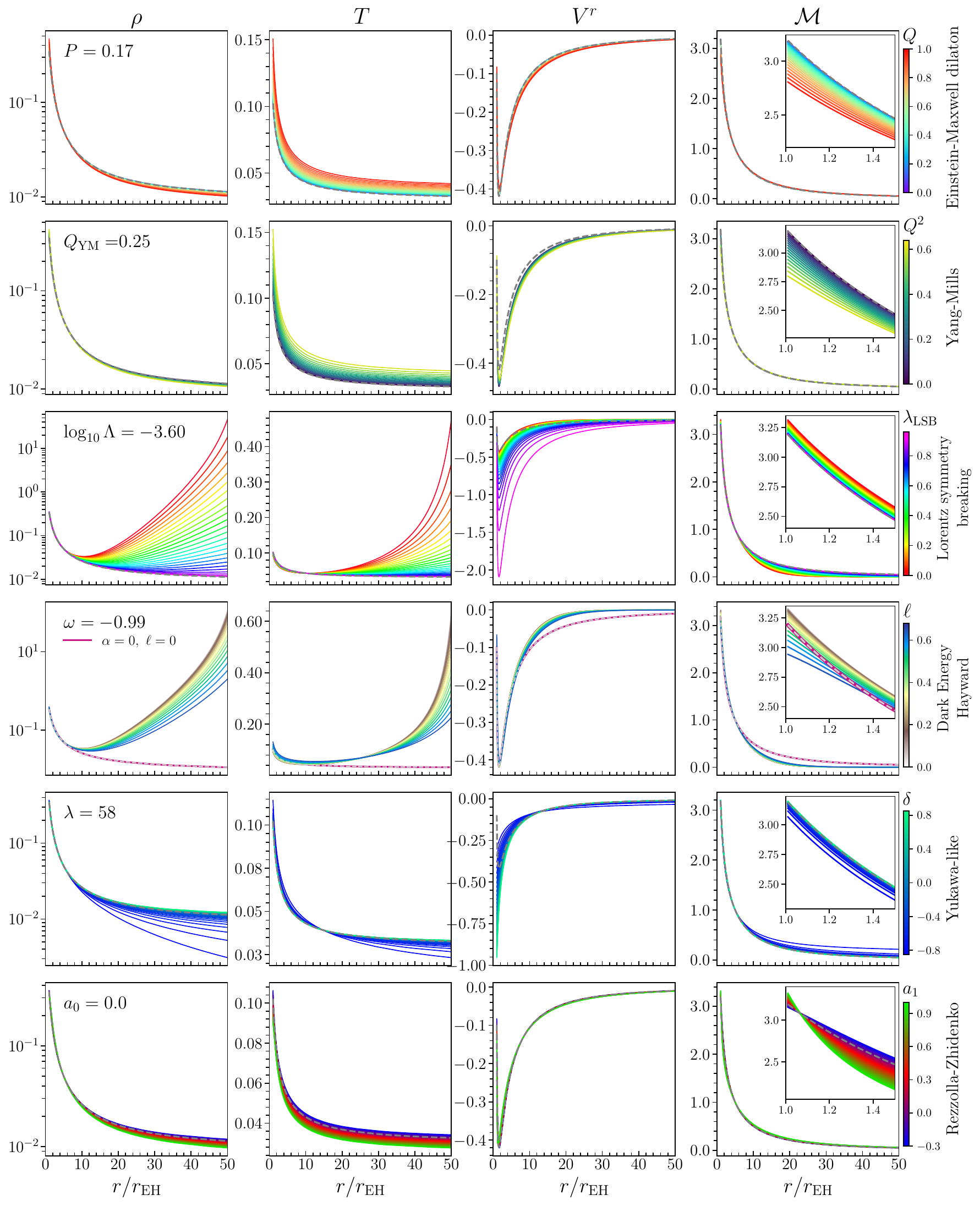}
    \caption{Radial profiles of gas properties. The left and right columns show black hole spacetimes with one and two free parameters, respectively. From left to right, we show the rest-mass density 
    $\rho$, proton temperature $T = p/\rho$, radial velocity component $V^r$, and Mach number 
    $\mathcal{M} = V/c_s$. For visualization purposes, the radial coordinate is normalized by 
    the event horizon radius of the black hole, and the domain spans $r/r_{\rm EH} \in (1, 50]$. 
    Colored lines represent solutions for different values of the black hole parameters (see 
    Table \ref{tab:spaceparameters}), while the gray dashed line corresponds to the Schwarzschild 
    black hole solution. 
    For Janis-Newman-Winicour and SV wormhole spacetimes we normalise using pseudo-event 
    horizons of the form $r_{\rm EH}=2M/\gamma$ and $r_{\rm EH}=2M-0.39\tilde a$, respectively.}
    \label{fig:1Dplots}
    \end{figure*}

\begin{figure*}
    \includegraphics[width=0.453\textwidth]{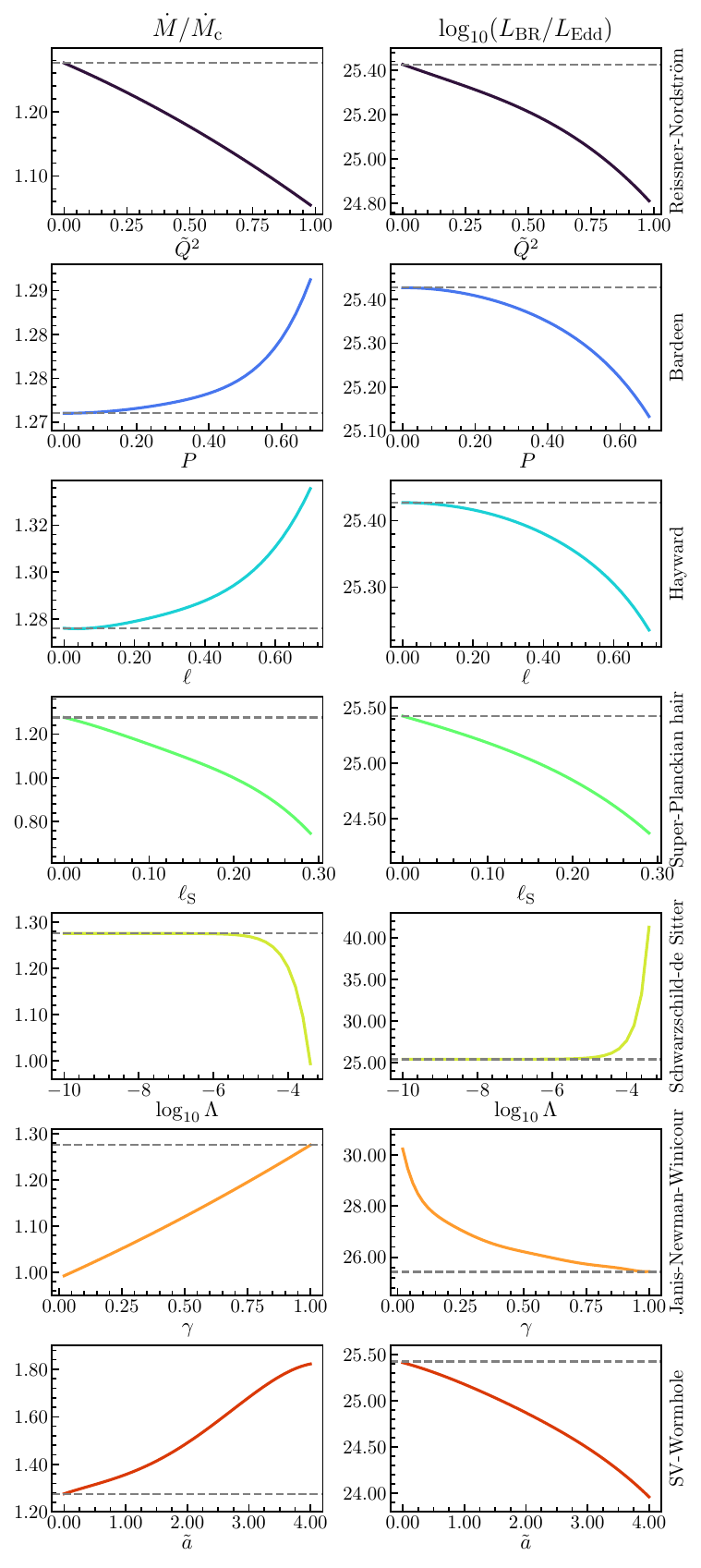}
    \includegraphics[width=0.46\textwidth]{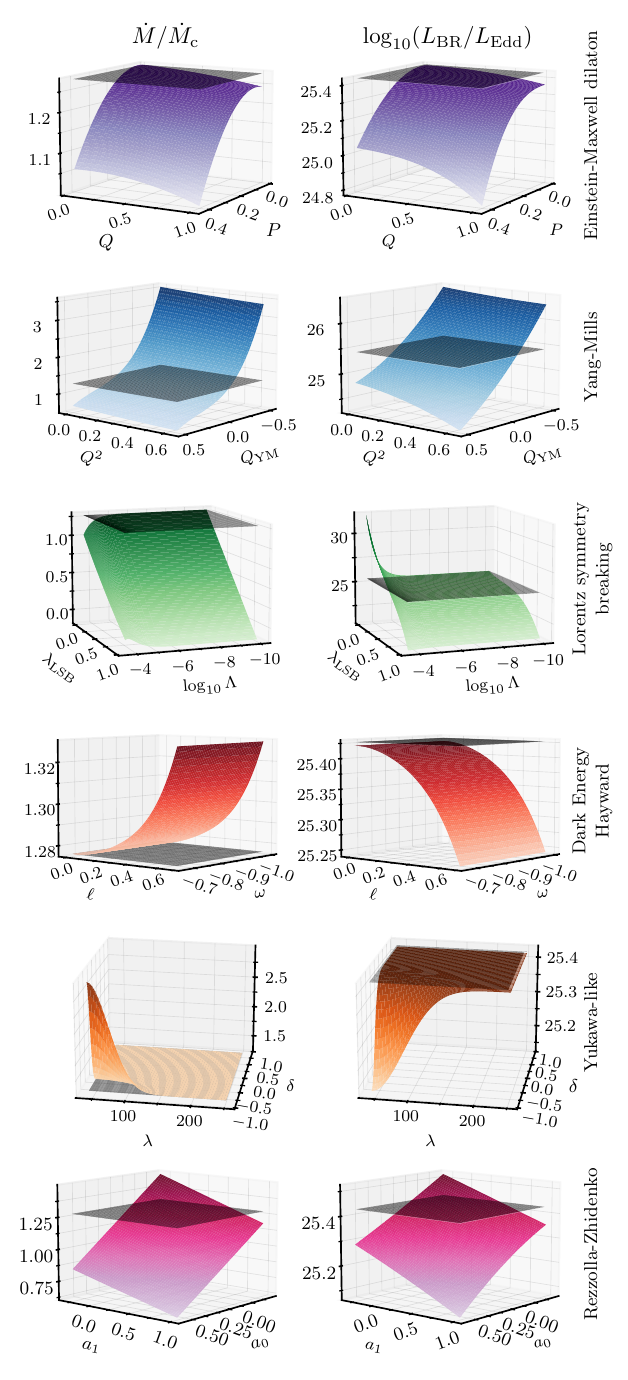}
    \caption{Normalized mass accretion rate, $\dot{M}/\dot{M}{\rm c}$, and 
    bremsstrahlung luminosity, $L_{\rm BR}/L_{\rm Edd}$, computed from 
    Eqs. \eqref{eq:mref} and \eqref{eq:Luminosity}. The left and right columns 
    show black hole spacetimes with one and two free parameters, respectively. 
    The fitting formulas are listed in Table \ref{tab:Mdot_L_fittings}. The gray dashed line (left) and gray surface (right) denote the Schwarzschild limit. Here, $\dot{M}{\rm c}$ is the critical mass accretion rate and $L_{\rm Edd}$ is the Eddington luminosity. 
    }
    \label{fig:Mdot_L_fits}
    \end{figure*}

    In this appendix, we present the complete set of Michel solutions for the fourteen black hole 
    spacetimes considered in this work. Figure \ref{fig:1Dplots} displays the rest-mass density, 
    temperature, radial velocity, and Mach number obtained by solving Eq. \eqref{eq:generaleq} 
    for the different parameter choices of each spacetime. Figure \ref{fig:Mdot_L_fits}  shows 
    the corresponding normalized mass accretion rates and bremsstrahlung luminosities.

    \end{document}